\documentclass[prb,aps,twocolumn,nofootinbib,showpacs,superscriptaddress,floatfix]{revtex4}
\usepackage{graphicx}
\usepackage{amsfonts}
\usepackage{amsbsy,bm,amssymb,amsmath}
\usepackage{dcolumn}
\usepackage{bm}
\usepackage{epsfig}
\usepackage{color}
\newcommand{\bra}[1]{\langle#1|}

\newcommand{\ket}[1]{|#1\rangle}

\begin{document}
\title{Preserving entanglement and nonlocality in solid-state qubits by dynamical decoupling}

\author{R. Lo Franco}
\email{rosario.lofranco@unipa.it}
\affiliation{Dipartimento di Fisica e Chimica, Universit\`{a} degli Studi di Palermo, Via Archirafi 36, 90123 Palermo, Italy}
\affiliation{Instituto de F{\'{i}}sica de S{\~{a}}o Carlos, Universidade de S{\~{a}}o Paulo, Caixa Postal 369, 13560-970 S{\~{a}}o Carlos, S{\~{a}}o Paulo, Brazil}
\affiliation{School of Mathematical Sciences, The University of Nottingham, University Park, Nottingham NG7 2RD, United Kingdom}
\author{A. D'Arrigo}
\affiliation{Dipartimento di Fisica e Astronomia,
Universit\`a di Catania, Via Santa Sofia 64, 95123 Catania, Italy}
\affiliation{CNR-IMM  UOS Catania (Universit\`a), 
Consiglio Nazionale delle Ricerche, Via Santa Sofia 64, 95123 Catania, Italy}
\author{G. Falci}
\affiliation{Dipartimento di Fisica e Astronomia,
Universit\`a di Catania, Via Santa Sofia 64, 95123 Catania, Italy}
\affiliation{CNR-IMM  UOS Catania (Universit\`a), 
Consiglio Nazionale delle Ricerche, Via Santa Sofia 64, 95123 Catania, Italy}
\affiliation{Istituto Nazionale di Fisica Nucleare, Sezione di Catania, Via Santa Sofia 64, 95123 Catania, Italy}
\author{G. Compagno}
\affiliation{Dipartimento di Fisica e Chimica, Universit\`{a} degli Studi di Palermo, Via Archirafi 36, 90123 Palermo, Italy}
\author{E. Paladino}
\affiliation{Dipartimento di Fisica e Astronomia,
Universit\`a di Catania, Via Santa Sofia 64, 95123 Catania, Italy}
\affiliation{CNR-IMM  UOS Catania (Universit\`a), 
Consiglio Nazionale delle Ricerche, Via Santa Sofia 64, 95123 Catania, Italy}
\affiliation{Istituto Nazionale di Fisica Nucleare, Sezione di Catania, Via Santa Sofia 64, 95123 Catania, Italy}

\begin{abstract}
In this paper we study how to preserve entanglement and nonlocality under dephasing produced by classical noise with large 
low-frequency components, as $1/f$ noise, by Dynamical 
Decoupling techniques. 
We first show that quantifiers of entanglement and nonlocality 
satisfy a closed relation valid for two independent qubits 
locally coupled to a generic environment under pure 
dephasing and starting from a general class of initial states. 
This result allows to assess the efficiency of 
pulse-based dynamical decoupling for protecting 
nonlocal quantum correlations between
two qubits subject to pure-dephasing local 
random telegraph and $1/f$-noise. 
We investigate the efficiency of an ``entanglement memory''
element under two-pulse echo and under sequences of   
periodic, Carr-Purcell and Uhrig dynamical decoupling. 
The Carr-Purcell sequence is shown to outperform the other
sequences in preserving entanglement against both random 
telegraph and $1/f$ noise. 
For typical $1/f$ flux-noise figures in superconducting nanocircuits, we show that entanglement and its nonlocal features 
can be efficiently stored up to times
one order of magnitude longer than natural entanglement 
disappearance times 
employing pulse timings of current experimental reach.
\end{abstract}

\date{\today}


\maketitle

\section{Introduction}
Controlling the dynamics of entanglement and preventing its 
disappearance due to decoherence~\cite{kr:203-zurek-rmp-decoherence} and via peculiar phenomena as 
the entanglement sudden death (ESD)~\cite{yu2009Science},
is a key requisite for any implementation of quantum 
information processing. 
For instance an entanglement memory element based on 
solid-state qubits will be strongly affected by dephasing 
due to noise sources with typical $1/f$ power 
spectrum~\cite{PaladinoReview2014}. 
To circumvent this problem it has been proposed to use 
hybrid systems combining superconducting nanocircuits 
with microscopic systems (atoms or defects), these latter  
having much longer coherence times and being 
suited to store quantum 
information~\cite{norireview,obrienreview}. 
Actually networking with different platforms 
has a much wider scenario of potential applications, 
and it is believed 
to be the pathway towards the implementation 
of quantum hardware, despite of the obvious advantages 
(fabrication, control and scalability) of 
performing both quantum operations and storage 
on a single platform. 
Such applications, and other 
technologies as security-proof quantum key distribution and quantum communication complexity~\cite{acin2006PRL,gisin2007natphoton,pironio2010Nature,lucamarini2012PRA}, 
depend critically on the existence of 
quantum correlations {\em and} nonlocality,  
witnessing non-classically-reproducible entanglement~\cite{horodecki2009RMP,bellomo2008bell}.

In this paper we address the relevant and still unsolved  
question of how to preserve {\em entanglement and nonlocality} 
under dephasing produced by classical noise with large 
low-frequency components, as $1/f$ noise. To this end we 
investigate protection by dynamical decoupling (DD) 
techniques~\cite{viola1998PRA,PaladinoReview2014} 
focusing on an ``entanglement memory'', 
physically implemented by a bipartite solid state 
nanodevice. The physical message of our work is that 
DD operated by control resources within the present technologies allows to preserve quantum correlations for times long enough to perform two-qubit 
quantum operations.

Originally developed 
in nuclear magnetic resonance~\cite{vandersypen2005RMP} 
DD techniques find important applications 
to quantum hardware~\cite{wiseman2010book}. They  
are open-loop (feedback-free) control methods for 
Hamiltonian engineering, 
thereby they do not require 
additional resources as encoding overheads 
or measurements capabilities.
The strategy is dynamical averaging of environmental 
noise by suitably tailored pulse 
sequences~\cite{viola1998PRA}. The prototype is 
spin-echo~\cite{Mims1972}, employing a single $\pi$ pulse 
to cancel unwanted static couplings in the Hamiltonian,
since the effect of the environment accumulated before the 
pulse is canceled during the subsequent ``reversed'' evolution. 
The Periodic DD (PDD), consisting in a  
train of such pulses separated by $\Delta t$,  
attenuates the effects of noise~\cite{viola1998PRA,viola1999PRL}
especially at low frequencies, 
$\omega \lesssim 1/\Delta  t<1/\tau_c$, where 
$\tau_c$ is the correlation time of the environment.
In this work we consider PDD along with improved versions 
of DD sequences, 
namely the Carr-Purcell (CP)~\cite{carr1954PhysRev} and the 
Uhrig DD (UDD)~\cite{uhrig2007PRL,yang2008PRL}
sequences. Pulse timing in these latter protocols 
is arranged in a way to produce higher order 
cancellations~\cite{biercuk2011JPB} 
in the Magnus expansion of the system 
``average Hamiltonian''~\cite{vandersypen2005RMP}, 
yielding a stronger protection from noise.

It is known that DD techniques efficiently fight 
decoherence~\cite{viola1998PRA} affecting single qubits, 
especially in the relevant case 
of the $1/f$ environment\cite{PaladinoReview2014}. 
Indeed it has been shown that PDD achieves substantial 
decoupling, mitigating dephasing due to random 
telegraph noise (RTN) and to $1/f$ noise, both 
for quantum~\cite{falci2004PRA,Rebentrost2009,lutchyn2008PRB} 
and for classical~\cite{faoro2004PRL,bergli2007PRB,Cheng2008,Gutmann2005,cywinsky2008PRB} models. More recently 
the performances of optimized sequences have been 
analyzed~\cite{biercuk2011JPB,du2009nature,Lee2008}.  
Routinely in experiments with superconducting qubits
spin- or 
Hahn-echo~\cite{nakamura2002PRL,Bertet2005,ithier2005PRB,Yoshihara2006} are seen to reduce 
defocusing due to noise sources 
of various origin with $1/f^\alpha$ spectrum. 
Recently control by PDD, CP, 
CP-Meiboom-Gill and UDD sequences has been successfully 
implemented~\cite{bylander2011NatPhys,yan2012PRB,Yuge2011PRL}.

The possibility to preserve entanglement via various 
DD sequences has been also theoretically 
investigated recently~\cite{muhktar2010PRA1,muhktar2010PRA2,wang2011PRA,pan2011JPB} for finite-dimensional or harmonic 
quantum environments.
Concerning $1/f$ noise,  
enhancement of the lifetime of an 
entangled state of a superconducting flux qubit coupled 
to a microscopic two-level 
fluctuator~\cite{gustavsson2012PRL}
has been observed under DD sequences. 
Experimental demonstrations of DD protection  
of bipartite entanglement from a solid-state environment 
have also been reported~\cite{wang2007PRL,roy2011PRA,shulman2012Science,dolde2013NatPhys} 
for ensembles of nuclear, impurity and electron spin-$1/2$ .

The results we present in this paper show 
that DD sequences are able to preserve entanglement, ensuring at 
the same time the existence of nonlocality, for a wide class 
of mixed initial states in a pure dephasing environment.
To this end we proof a relation between entanglement quantified 
by the concurrence~\cite{wootters1998PRL}, and nonlocality 
identified by the violation of a Bell 
inequality~\cite{horodecki2009RMP}.  
For realistic figures of $1/f$ 
noise~\cite{bylander2011NatPhys,gustavsson2012PRL} protection 
for times more than one order of magnitude longer than ESD times
is achieved, allowing advanced applications based on nonlocality.
Notice that  in our proposal DD fighting 
$1/f$ noise is implemented avoiding non-local control, and
using pulse rates well within present experimental
capabilities~\cite{bylander2011NatPhys,gustavsson2012PRL}.

The paper is organized as follows. In Section II we derive the 
relation between entanglement and nonlocality for 
extended Werner-like (EWL) under a pure dephasing dynamics. 
In Section III we introduce the model and the DD sequences, 
illustrating then our approach to evaluate the
concurrence. In Section IV we analyze the case study of RTN.
We address the performance of DD sequences in the 
presence of local pure dephasing $1/f$ noise in Section V.
Finally Section VI is devoted to the conclusions.

\section{Relation between entanglement and nonlocality at pure dephasing}
Strongly entangled systems are characterized by the 
presence of quantum correlations that cannot be reproduced 
by any classical local model. In these cases Quantum Mechanics
exhibits nonlocality, which would guarantee resources for 
quantum technologies as secure quantum cryptography~\cite{acin2006PRL,gisin2007natphoton,lucamarini2012PRA}. 
For pure states entanglement always 
corresponds to the presence of nonlocality, but 
this is not the case in general. 
In fact mixed states exist whose correlations 
can be reproduced by a classical local 
model~\cite{horodecki1995PLA}, while they are entangled, 
as indicated by a nonzero value of the 
concurrence~\cite{wootters1998PRL} $C(t)$. 
Nonlocality in such cases is unambiguously identified if 
Bell inequalities are violated. Therefore 
the Bell function $\mathcal{B}$, as defined by the 
Clauser-Horne-Shimony-Holt (CHSH) form~\cite{horodecki2009RMP}, 
can be used to seek whether the system exhibits nonlocal correlations, which occurs with certainty if $\mathcal{B}>2$.

The possible existence of closed relations between quantifiers of entanglement and nonlocality is currently an open issue
of special interest in dynamical contexts~\cite{mazzolapalermo2010PRA,horst2013PRA,bartkiewicz2013PRA}. 
A relevant question is establishing, for a given 
time evolution, whether a threshold value of 
concurrence exists ensuring nonlocal quantum correlations. 
More generally, 
the presence of such correlations would guarantee resources for 
quantum technologies as secure quantum cryptography \cite{acin2006PRL,gisin2007natphoton,lucamarini2012PRA}, thereby 
efficient DD sequences must preserve entanglement above this 
threshold.

In this Section we analyze the relation between quantifiers of entanglement 
and nonlocality for two noninteracting qubits, $A$ and $B$,
locally subject to a pure dephasing interaction with the environment. 
Each qubit has Hamiltonian
($\hbar=1$, $s=A,B$)
\begin{equation}\label{pure-dephasingH}
H_s= -{\Omega_s \over 2}\sigma_z^s  - {\hat{X}^s \over 2} \sigma_z^s + \hat{H}_R^s,
\end{equation}
where $\Omega_s$ is the Bohr frequency of qubit-$s$ and $\hat{X}^s$ represents a collective environmental operator
coupled to the same qubit. The free evolution of the environment is included in $\hat{H}_R^s$. The overall Hamiltonian is 
thus $H=H_A+H_B$. Results of the present Section are valid for any $\hat{H}_R^s$ and $\hat{X}^s$. 

We suppose the two qubits are prepared in an EWL state
\begin{equation}\label{EWLstates}
   \rho_1=r \ket{1_{a}}\bra{1_{a}}+\frac{1-r}{4}\openone_4,\quad
    \rho_2=r \ket{2_{a}}\bra{2_{a}}+\frac{1-r}{4}\openone_4,
\end{equation}
where the pure parts $\ket{1_{a}}=a\ket{01}+b\ket{10}$ and $\ket{2_{a}}=a\ket{00}+b\ket{11}$ are, respectively, 
the one-excitation and two-excitation Bell-like states with $|a|^2+|b|^2=1$. When $a=b=1/\sqrt{2}$ the EWL states reduce to
the Werner states, a subclass of Bell-diagonal states \cite{verstraete2002PRL,horodecki2009RMP}. The density matrix of EWL
states, in the computational basis $\{\ket{0}\equiv\ket{00},\ket{1}\equiv\ket{01}, \ket{2}\equiv\ket{10}, 
\ket{3}\equiv\ket{11}\}$, is non-vanishing only along the diagonal and anti-diagonal (X form). The purity 
$P=\mathrm{Tr}(\rho^2)$ of EWL states only depends on the purity parameter $r$ and it is given by $P=(1+3r^2)/4$. 
The initial entanglement is equal for both the EWL states of Eq.~(\ref{EWLstates}) with concurrence 
$C_{\rho_1}(0)=C_{\rho_2}(0)=2\mathrm{max}\{0,(|ab|+1/4)r-1/4\}$. Initial states are thus entangled for $r>\bar{r}=(1+4|ab|)^{-1}$. 

Since the two qubits are noninteracting, the evolution of entanglement and nonlocality can be simply obtained from the
knowledge of single-qubit dynamics \cite{bellomo2007PRL}.
Under a pure dephasing evolution, for each qubit the diagonal elements of the density matrix in the eigenstate basis
remain unchanged. The single-qubit coherences evolve in time $q_s(t)\equiv\rho_{01}^s(t)/\rho_{01}^s(0)$, 
the explicit time dependence being specified by the environmental properties and the interaction term. 
If the system is subject to pure dephasing only,  the X form of 
the density matrix is kept at $t>0$. 
In particular, diagonal elements remain constant whereas 
antidiagonal elements evolve in time. They 
are related to the single qubit coherences by 
$\rho_{12}(t)=\rho_{12}(0) q_A(t) q_B^\ast(t)$ for the initial state $\rho_1$ and 
$\rho_{03}(t)=\rho_{03}(0) q_A(t) q_B(t)$ for $\rho_2$.
The concurrences at time $t$ for the two initial states of Eq.~(\ref{EWLstates}) are given by \cite{yu2007QIC} 
$C_{\rho_1}(t)=2\mathrm{max}\{0,|\rho_{12}(t)|-\sqrt{\rho_{00}(0)\rho_{33}(0)}\}$ and 
$C_{\rho_2}(t)=2\mathrm{max}\{0,|\rho_{03}(t)|-\sqrt{\rho_{11}(0)\rho_{22}(0)}\}$.
For the pure dephasing evolution, it easy to show that 
$C_{\rho_1}(t)=C_{\rho_2}(t) \equiv C(t)$ with
\begin{equation}\label{Kt}
C(t)=2\mathrm{max}\{0,r|a|\sqrt{1-|a|^2}|q_A(t)q_B(t)|-(1-r)/4\}.
\end{equation}
We now turn to nonlocality. 
The maximum CHSH-Bell function $\mathcal{B}$ for a general X state can be found in
analytic form \cite{horodecki1995PLA}. It can be expressed as $\mathcal{B}=\mathrm{max}\{\mathcal{B}_1,\mathcal{B}_2\}$, where 
$\mathcal{B}_1$, $\mathcal{B}_2$ are functions of the density matrix elements \cite{bellomo2010PLA,mazzolapalermo2010PRA}. This 
quantity has been studied for independent qubits each coupled to a bosonic reservoir 
(cavity) with Markovian \cite{miran2004PLA} and non-Markovian \cite{bellomo2008bell,bellomo2008nonlocal} features. 
For independent qubits subject to local pure dephasing noise, the two functions $\mathcal{B}_1$, $\mathcal{B}_2$ have the same 
form for the initial EWL states of Eq.~(\ref{EWLstates}) and are given by
\begin{eqnarray}\label{B1B2}
&\mathcal{B}_1(t)=2\sqrt{r^2+4r^2|a|^2(1-|a|^2)|q_A(t) q_B(t)|^2},&\nonumber\\
&\mathcal{B}_2(t)=4\sqrt{2}r|a|\sqrt{1-|a|^2}|q_A(t) q_B(t)|.&
\end{eqnarray}
It is easily seen that $\mathcal{B}_1(t)$ is always larger than or equal to $\mathcal{B}_2(t)$, so that the maximum Bell function 
is $\mathcal{B}(t)=\mathcal{B}_1(t)$. 

To find a closed relation between $\mathcal{B}(t)$ and $C(t)$ we first observe that, in order to achieve nonlocality, 
two-qubit entanglement is necessary, i.e. $C(t)>0$. 
Under these conditions 
$C(t)=2[r|a|\sqrt{1-|a|^2}|q_A(t)q_B(t)|-(1-r)/4]$ and from Eq. (\ref{B1B2}) we obtain 
\begin{equation}\label{BversusC}
\mathcal{B}(t)=2\sqrt{r^2+4[C(t)/2+(1-r)/4]^2}.
\end{equation}
We remark that this result is valid for {\em any} local pure-dephasing qubit-environment interaction, starting from 
an initial EWL state with a generic value of 
$a\neq 0,1$.
For example, when $r=1$ (initial Bell-like state), Eq.~(\ref{BversusC}) reduces to $\mathcal{B}(t)=2\sqrt{1+C(t)^2}$.
This relation, known when the system is in a pure state \cite{gisin1991PLA} or in a Bell-diagonal 
state \cite{verstraete2002PRL}, is here found to persist {\em during} the system evolution for more general states.

The threshold value for $C(t)$ ensuring that at time $t$ it is $\mathcal{B}(t)>2$ immediately derives   
from Eq.~(\ref{BversusC})
\begin{equation}\label{Cthreshold}
C_\mathrm{th}=\sqrt{1-r^2}-(1-r)/2.
\end{equation}
Thus, for initial EWL states evolving under any pure dephasing interaction, the system exhibits nonlocality
at time $t$ provided the concurrence $C(t)$ is larger than a threshold value $C_\mathrm{th}$ 
depending {\em only} on the system initial purity. The threshold is a decreasing function of the purity and for $r=1$ it is $C_\mathrm{th}=0$. 

This result has relevant implications in those quantum computing platforms allowing for accurate initial
state preparation.
In particular this is the case of superconducting nanodevices.
Preparation of entangled states has been recently implemented in different laboratories 
\cite{schoelkopf2009Nature,Dicarlo2010,Neeley2010,Lucero2012,Mariantoni2011,Fedorov2012,Reed2012,Chow2012,Rigetti2012}.
For instance, entangled states of two superconducting qubits with purity 
$\approx 0.87$ and fidelity to ideal Bell states $\approx 0.90$ have been experimentally generated by using 
a two-qubit interaction, mediated by a cavity bus in a circuit quantum electrodynamics architecture \cite{schoelkopf2009Nature} . 
These states may be approximately described as EWL states with $r=r_\mathrm{exp}\approx0.91$ and 
$|a|=1/\sqrt{2}$, giving a value of initial concurrence 
$C=0.865$ and a threshold value for having nonlocality 
with certainty $C_\mathrm{th}\approx0.37$.
In the reminder of this paper, except when explicitly 
mentioned, we will use these parameters for the initial 
EWL state, and the threshold value $C_\mathrm{th}\approx0.37$ 
as a benchmark for entanglement protection. 

\section{Model and dynamical decoupling sequences}

We consider a two-qubit entanglement memory element where each qubit is locally subject
to an ensemble of classical bistable fluctuators at pure dephasing 
and to pulse-based DD as modeled by
\begin{equation}\label{HDD}
H_s^\mathrm{DD}=H_s+\mathcal{V}_s(t),
\end{equation}
where  single qubit Hamiltonian $H_s$ is of the form of Eq. (\ref{pure-dephasingH}) and quantum control is operated by the external 
field included in $\mathcal{V}_s(t)$. 
The environmental operator $\hat X_s$  is here replaced by  the stochastic process $X_s(t)=\sum_i^{N} v_i \xi_i(t)$ where
$\xi_i(t)$ is a bistable symmetric process randomly switching between $0$ and $1$ with an overall rate $\gamma_i$.
The power spectrum of the equilibrium fluctuations of each $v_i\xi_i(t)$,
\begin{equation}
S_i(\omega) = \int_{-\infty }^\infty d t \; v_i^2 \, [\langle \xi_i(t) \xi_i(0) \rangle - \langle \xi_i(t)^2 \rangle]\, e^{i \omega t}
\label{def_spectrum}
\end{equation}
is a Lorentzian $S_i(\omega)=v_i^2\gamma_i/[2(\gamma_i^2+\omega^2)]$.
Following the standard procedure~\cite{weissman1988RMP,PaladinoReview2014}, we model $1/f$ noise as due 
to an ensemble of $N$ random telegraph processes,
individual rates being distributed in the interval
$\gamma_i \in [\gamma_m, \gamma_M]$ with probability density 
$\propto 1/\gamma$. This yields the power spectrum
\begin{equation}\label{eq:1overf-powerspec}
S^{1/f}(\omega) \approx 
{\pi\sigma^2 \over \ln(\gamma_M/\gamma_m) \,\omega}
\end{equation}
where the noise variance $\sigma$ is related to 
the distribution of couplings $v$. Assuming  
a narrow distribution about the average $\bar{v}$
we have $\sigma^2=\bar{v}^2N/4$~\cite{paladino2002PRL}.

For DD 
we consider sequences of an {\em even} number $n$ of instantaneous $\pi$-pulses 
about the $x$-axis, orthogonal to the qubit-environment interaction.
The pulses are applied at times 
$t_k=\delta_k t$, where $t$ is the total evolution time and 
$0\leq\delta_k\leq1$ with $k=1,\ldots,n$. In PDD 
$\delta_k=k/n$ and $\Delta t=t/n$, the last pulse being 
applied at the observation time $t$.  
A PDD sequence with $n=2$ corresponds
to the echo procedure. In the CP sequences 
$\delta_k =(k-1/2)/n$, while in UDD  
$\delta_k=\sin^2[\pi k/(2n+2)]$. 
In the limit of a two-pulse cycle, $n=2$, UDD reduces to the 
CP sequence.

We suppose the qubits are prepared at time $t=0$ in a EWL state by some interaction which is thereafter switched off.
Since both noise and decoupling sequences act locally, the two-qubit density matrix is entirely expressed by 
the single qubit coherences $q_s(t)$. For the PDD we will rely on the exact analytic expression
for a qubit affected by a {\em quantum} environment of bistable impurities \cite{falci2004PRA,falci2005PhysE}. Here we specialize to the classical
limit where each impurity produces RTN and compare with the Gaussian approximation where the coherence
can be expressed as $q_s(t) = \exp{\{- \Gamma_s(t)}\}$ with \cite{biercuk2011JPB} 
\begin{equation}\label{Gausscoherence}
\Gamma_s(t)=\int_0^\infty \mathrm{d}\omega S_s(\omega) \frac{f(\omega t)}{\pi \omega^2},
\end{equation}
where the ``filter function'' $f(\omega t)$
is specific to the pulse sequence~\cite{uhrig2007PRL}. 
For PDD  it reads~\cite{cywinsky2008PRB} 
$f_{\rm{PDD}}(\omega t) = 
2 \tan^2[\omega t/(2n)] \sin^2(\omega t/2)$.
For the CP and UDD we will rely on the analysis of 
Ref.~\cite{cywinsky2008PRB} where it has been found 
for single qubit coherence that down to a relatively small 
pulse rate the effect of RTN is reasonably approximated by 
a Gaussian (Ornstein-Uhlenbeck) process even under strong 
coupling conditions (see next Section
for a quantitative definition). Therefore for CP and UDD 
we will resort to the Gaussian approximation 
Eq.~(\ref{Gausscoherence})
with filter functions (for an even number of pulses $n$) are
$f_{\rm{CP}}(\omega t) = 8 \sin^4[\omega t/(4n)] \sin^2(\omega t/2)/\cos^2[\omega t/(2n)]$  and
$f_{\rm{UDD}}(\omega t) = \frac{1}{2} |\sum_{k=-n-1}^n 
(-1)^k \exp [i\frac{\omega t}{2}\cos\frac{\pi k}{n+1}]|^2$ 
respectively. 
Equipped with these expressions for the single qubit 
coherences we can investigate time evolution of 
the entanglement by Eq.~(\ref{Kt}), using 
$q_s(t)$ as given by the specific expression for each 
sequence.

\begin{figure}
\begin{center}
{\includegraphics[width=0.45\textwidth]{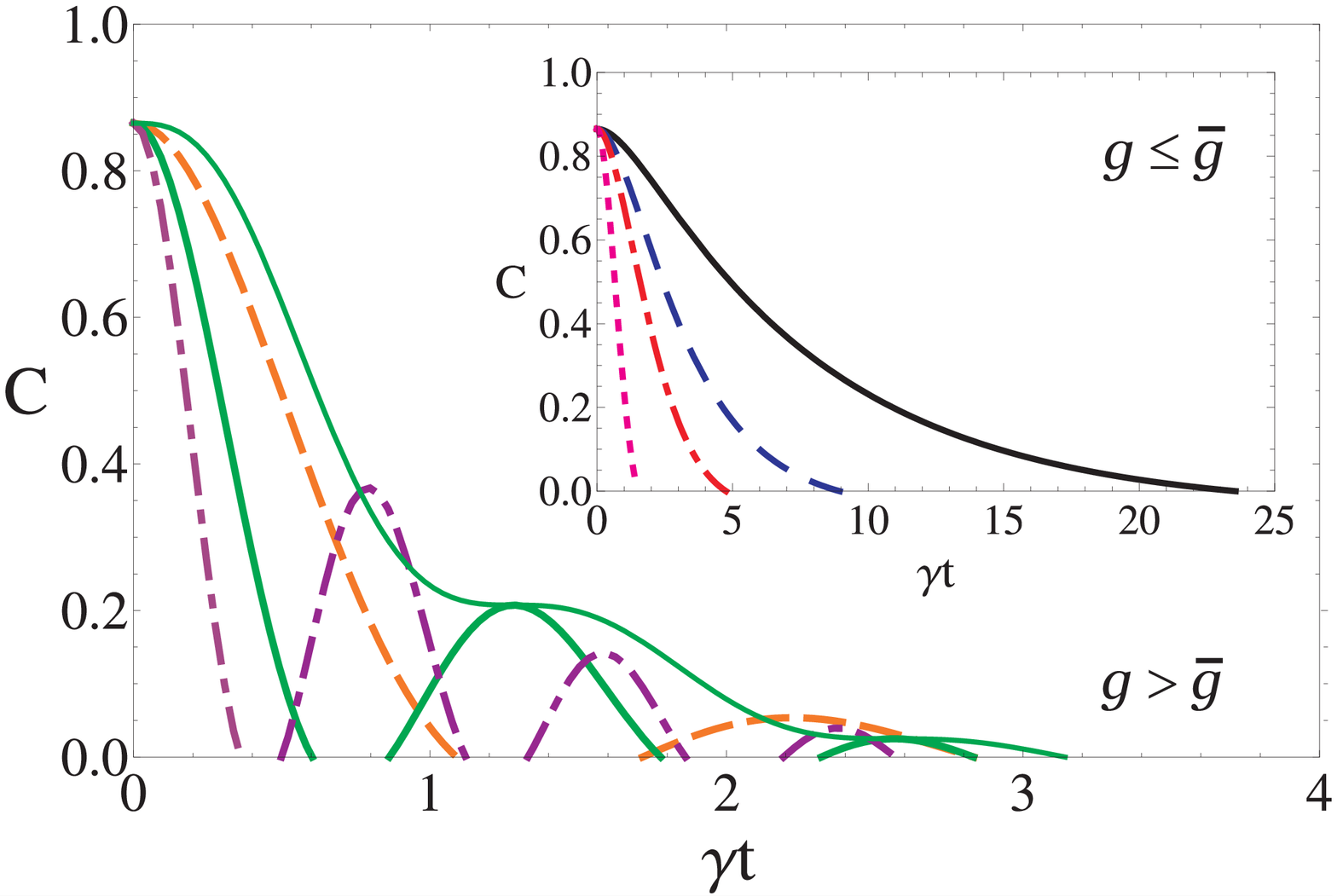}\vspace{0.5 cm}
\includegraphics[width=0.45\textwidth]{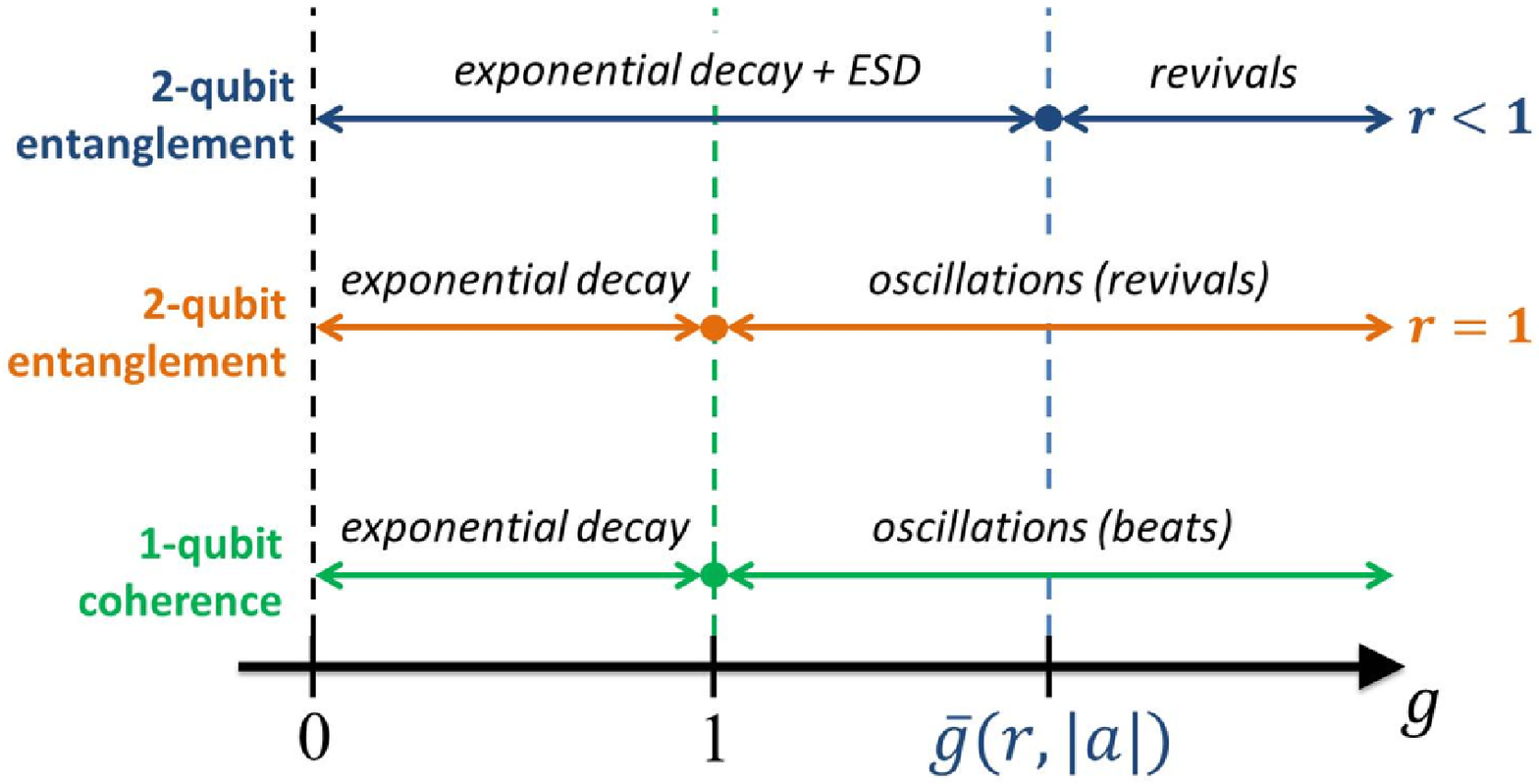}}
\caption{\label{fig:CnoBB}\footnotesize (Color online) 
Top panel: Concurrence $C$ as 
a function of $\gamma t$ under local RTNs with $g > \bar g \approx 2.3$ (EWL initial states with
 $r=0.91$, $|a|=1/\sqrt{2}$):  $g=3$ (dashed orange), $g=5$ (thick green) 
and $g=8$ (dot-dashed purple). 
We fixed $\delta p_0=0$ except for the thin green line which corresponds to $\delta p_0=\pm1$ for 
$g=5$. 
In the inset $g < \bar g$: $g=0.5$ (thick black), $g=0.8$ (dashed blue), 
$g=1.1$ (dot-dashed red) and $g=\bar{g}=2.3$ (dotted magenta). 
Bottom panel: sketch of the threshold values of the dimensionless coupling parameter $g$ separating
dynamical regimes for single qubit and entanglement dynamics.
}
\end{center}
\end{figure}
    
\section{Dynamical decoupling of random telegraph noise}
We now consider an entanglement memory element where each qubit 
is locally subject to RTN and to DD sequences. 
We first review the effect of RTN on evolution of the 
entanglement for the initial  EWL states of 
Eq.~(\ref{EWLstates}) in the absence of 
pulses~\cite{zhou2010QIP,lofranco2012PhysScripta}. 
This preliminary analysis puts the basis for the investigation of entanglement preservation by DD.

{\em Entanglement under local RTN --} \;\;
The single qubit dynamics under these conditions has been investigated in several papers and here we 
briefly summarize the main findings.
The qubit coherence for qubit $s=A,B $ is given by \cite{paladino2002PRL,galperin2006PRL} 
\begin{equation}\label{singlequbitcoherence}
q_s^{\mathrm{RTN}}(t)=e^{-i \Omega_s t} [A_s e^{-\frac{\gamma_s(1-\alpha_s)t}{2}}
+(1-A_s)e^{-\frac{\gamma_s(1+\alpha_s)t}{2}}],
\end{equation}
here $A_s=\frac{1}{2\alpha_s}(1+\alpha_s-ig_s\delta p_{0,s})$ and $\alpha_s=\sqrt{1-g_s^2}$ are expressed in terms
of $g_s=v_s/\gamma_s$ and $\delta p_{0,s}$ is the initial population difference of the two states $\xi_s=0,1$.
Repeated measurements with fully thermalized fluctuators 
are described by $\delta p_{0,s}=0$, whereas other choices 
(e.g. $\delta p_{0,s}=\pm 1$) are appropriate for 
nonequilibrium conditions~\cite{paladino2002PRL}. 
For a weakly coupled fluctuator $g_s \ll 1$ the coherence decays exponentially with the 
rate $\Gamma= S_s(0)= v_s^2/(2 \gamma_s)$, which is the standard
golden rule result. Under these conditions the Gaussian 
approximation applies to the bistable process.
In the strong coupling regime $g_s \geq 1$, the system exhibits damped beatings and, for $g_s\gg 1$
the decay rate saturates to $\gamma_s$. In this regime the non Gaussian nature of the stochastic process is
clearly visible in the qubit evolution~\cite{paladino2002PRL,galperin2006PRL}.

The concurrence  of two uncoupled qubits each subject to a RTN process is readily found using Eq.~(\ref{Kt}) 
with $q_s(t)$ given by Eq.~(\ref{singlequbitcoherence}).
For a pure initial entangled state, $r=1$, $a\neq0$, entanglement reflects the single-qubit coherence 
qualitative behavior. 
The concurrence either decays exponentially if $g_s<1$ or displays damped beatings if at least one $g_s$ 
is larger than $1$.  
The regime $r <1$ instead reveals the new phenomenon 
of ESD, i.e. the 
concurrence~\cite{yu2009Science} 
vanishes abruptly at a certain time $t_{ESD}$, 
and show qualitative different entanglement behavior for  
different values of the dimensionless couplings $g_s$. 
For identical qubit-RTN coupling conditions 
(i.e. for both $g_s = g$) 
a threshold value exists, separating a regime of 
exponential entanglement decay or ESD from a regime where 
entanglement revivals occur.
This threshold value, not yet reported in the literature, 
only depends on the parametrization of the initial state and 
it is given by\footnote{The expression of $\bar{g}(r,|a|)$ of 
Eq.~(\ref{gfirstrevival}) is 
obtained as follows. Firstly, one finds the time $t_\mathrm{max}$ corresponding to the first maximum of $C(t)$ of 
Eq.~(\ref{Kt}), assuming $g>1$. Secondly, one looks for the values of $g$ such that $C(t_\mathrm{max})>0$, which in turn 
gives $g>\bar{g}(r,|a|)$.} 
\begin{equation}\label{gfirstrevival}
\bar{g}(r,|a|)=\sqrt{1+4\pi^2\left[\ln\left(\frac{4r|a|\sqrt{1-|a|^2}}{1-r}\right)\right]^{-2}}.
\end{equation}
Since $r<1$, it is $\bar{g}>1$; for instance for $r=0.91$ and $|a|=1/\sqrt{2}$, we get $\bar{g}\approx 2.3$. 
When $g\leq \bar{g}$ the system displays ESD, whereas for  $g>\bar{g}$  a ``final death'' (FD)
of entanglement takes place, i. e. a definitive disappearance of entanglement after revivals. 
These behaviors are illustrated in the top panel of Fig.~\ref{fig:CnoBB}.
The different dynamical regimes for single qubit and entanglement dynamics with respect to $\bar{g}$ 
are schematically illustrated in the bottom panel of Fig.~\ref{fig:CnoBB}. 

The ESD time for $g < 1$ can be simply derived from Eq.~(\ref{singlequbitcoherence}). 
When $g \ll 1$ and $\delta p_0=0$ the first term of Eq.~(\ref{singlequbitcoherence}) 
is much larger than the second one and, from Eq.~(\ref{Kt}), the ESD time (for equal $\gamma_s = \gamma$) is found as \cite{lofranco2012PhysScripta}
\begin{equation}
t_\mathrm{ESD}^{\mathrm{RTN}}=-\frac{2/\gamma}{1-\sqrt{1-g^2}}
\ln\left(\frac{\sqrt{\frac{(1-g^2)(1-r)}{r|a| \sqrt{1-|a|^2}}}}{1+\sqrt{1-g^2}}\right).
\end{equation}
We checked numerically that the above expression provides a reasonable approximation up to  $g=0.9$.
For larger values of $g$,  the numerical ESD and FD times are reported in Table \ref{tabletESD}. 
\begin{table}
\begin{center}
\footnotesize
{\begin{tabular}{|c|cccccccc|}\hline \hline
& & & & $g$ &  & & &  \\
\cline{2-9}  & 0.1&0.5&1.1&$\bar{g}=2.3$&3&5&10&30\\
\hline $\gamma t_\mathrm{ESD}$& 600& 23.60& 4.75 &1.50& 1.09 &0.6 &0.28 & 0.09 \\
\hline $\gamma t_\mathrm{FD}$& 600& 23.60& 4.75&1.50& 2.83& 2.83& 2.68 & 2.76\\
\hline\hline
\end{tabular}}
\caption{\label{tabletESD}\footnotesize Dimensionless ESD and FD times (scaled with $\gamma$) for 
different values of $g$ for local RTNs and initial EWL states with $r=r_\mathrm{exp}=0.91$ and 
$|a|=1/\sqrt{2}$. For $g>\bar{g}$, $t_\mathrm{ESD}$ has been identified as the first time at which
entanglement disappears, $C(t_\mathrm{ESD})=0$, thus $t_\mathrm{FD} > t_\mathrm{ESD}$.}
\end{center}
\end{table}
The dependence of the single-qubit coherence $q_s(t)$ on the initial population difference
$\delta p_{0,s}$ qualitatively affects the entanglement dynamics for $g>\bar{g}$, while it leaves
it practically unchanged for $g\leq \bar{g}$ \cite{lofranco2012PhysScripta}. 
In particular, when $g>\bar{g}$, the concurrence for $\delta p_0=\pm1$ does not exhibit revivals and it is always larger 
than for $\delta p_0=0$.
The final death time is also longer than for $\delta p_0=0$ (see Fig.~\ref{fig:CnoBB} for $g=5$).

In the following, we shall see that the existence of a threshold value for $g$ plays a role in the efficiency of the
DD procedure to prevent complete entanglement disappearance under local RTNs. Hereafter the value of the initial 
population difference of RTN is set to the thermal equilibrium value $\delta p_0=0$.

\subsection{Entanglement echo}
For the two-pulse echo the qubit coherence reads \cite{falci2004PRA,galperin2006PRL}
\begin{equation}\label{coherenceecho}
q_s^\mathrm{e}(t)=\frac{e^{-\frac{\gamma_s t}{2}}}{\alpha_s^2}\left[\frac{1+\alpha_s}{2}e^{\alpha_s\frac{\gamma_s t}{2}}+
\frac{1-\alpha_s}{2}e^{-\alpha_s\frac{\gamma_s t}{2}}-(1-\alpha_s^2)\right],
\end{equation}
where $t=2\Delta t$.
For the sake of simplicity, we assume that the two qubits are acted by simultaneous pulses 
applied at times $\Delta t$ and $2 \Delta t$. 
We consider two fluctuators with equal switching 
rates, $\gamma_s \equiv \gamma$, but differently coupled to the respective qubits $v_A \neq v_B$ in order to address 
different coupling regimes, $g_A \neq g_B$. The general outcome of this analysis
is that the echo preserves entanglement with a qualitative behavior critically sensitive to
the values of $\gamma \Delta t$ and $g_s$. 

\begin{figure}[t!]
\begin{center}
{\includegraphics[width=0.40\textwidth]{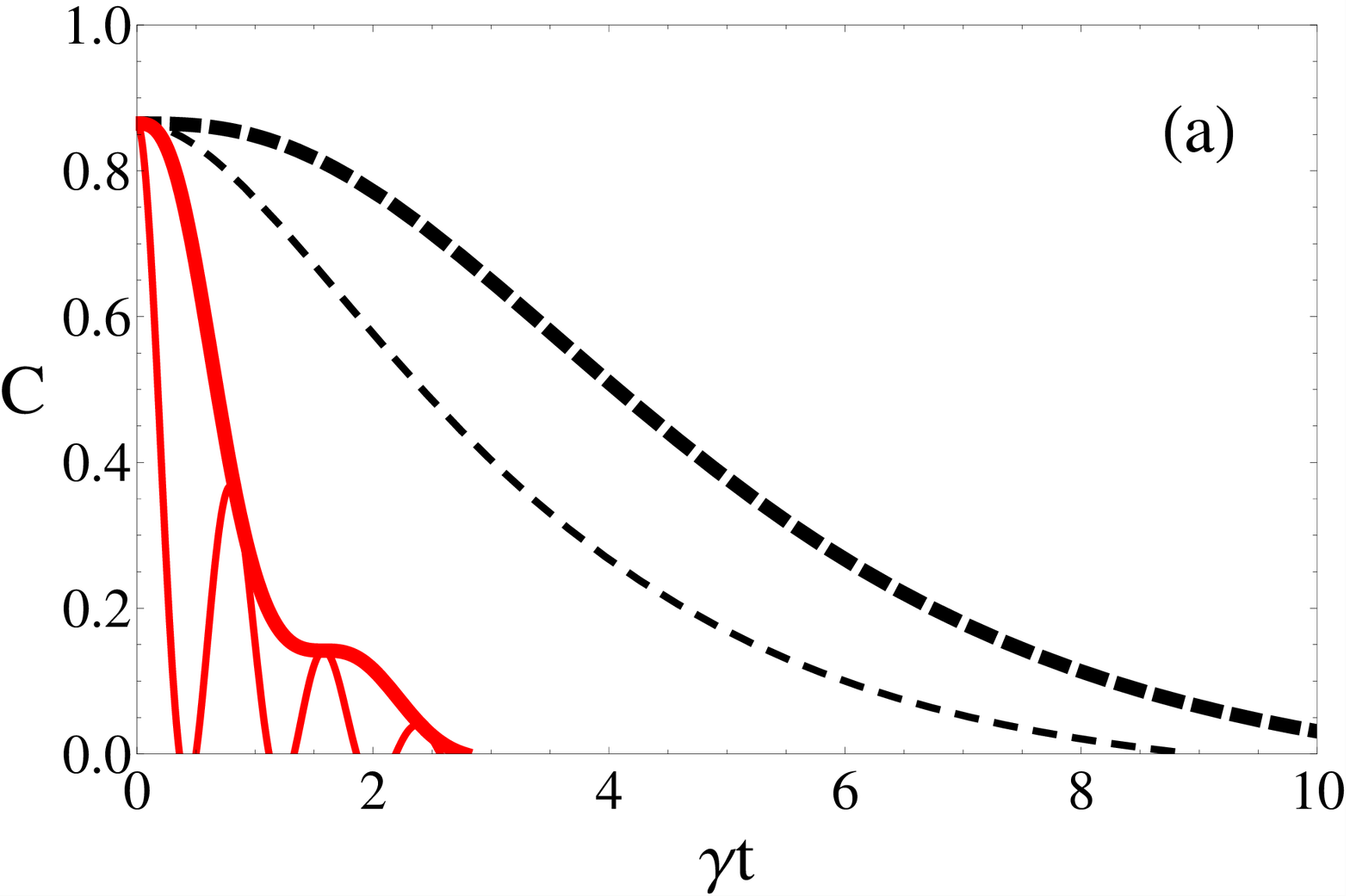}\vspace{0.5 cm}
\includegraphics[width=0.40\textwidth]{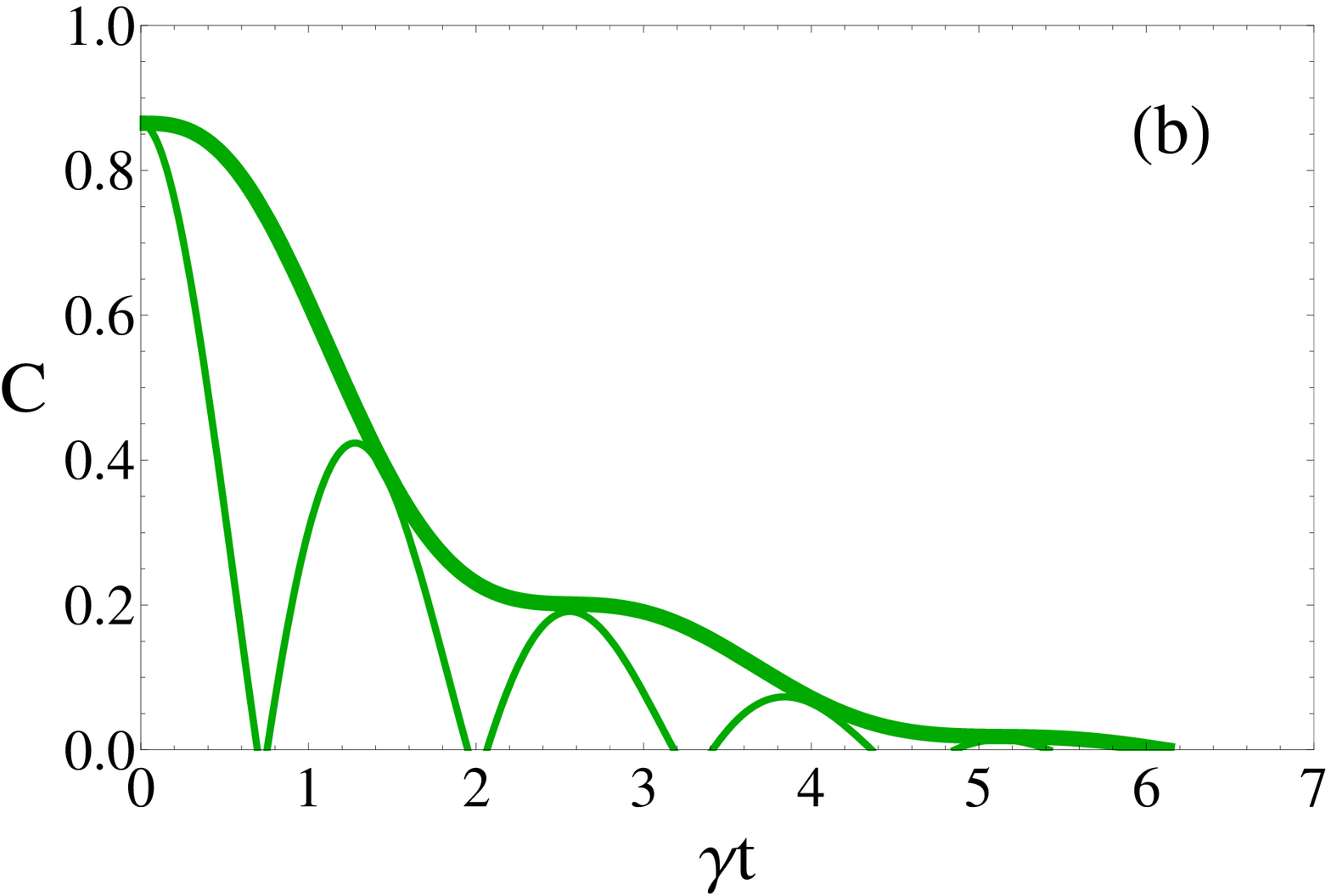}}
\caption{\label{fig:Cecho}\footnotesize 
(Color online) Concurrence $C(2\Delta t)$ at the end of the echo as a function of $\gamma t=2\gamma\Delta t$, showing 
different "entanglement echo" behavior depending on 
$g \gtrless \bar g= 2.3$ (for the chosen initial 
state mentioned in the text).  
Panel (a): Cases 
$g_s=g=0.8 <\bar{g}$ (dashed black lines) 
and $g=8 >\bar{g}$ (solid red lines), 
with echo (thick lines) and without (thin lines).
For $g=8$ notice the plateau at 
$\gamma \Delta t = 2 \pi/g \approx 0.8$. 
 Panel (b):  $g_A= 0.4$ and $g_B=5$, 
with echo (thick line) and without (thin line).
}
\end{center}
\end{figure}

When the qubits experience the same coupling conditions, 
$g_s \equiv g$, the entanglement-echo efficiency reflects 
the presence of ESD or of FD in the unconditioned evolution, 
i.e. it depends on whether $g$ is smaller or larger 
than $\bar{g}$.
This behavior of "entanglement echo" is 
illustrated in Fig.~\ref{fig:Cecho}(a). 
When $g\leq\bar{g}$ the ESD time is delayed, 
whereas for $g>\bar{g}$ the dynamical structure of entanglement 
revivals and dark periods is {washed} out
by the echo procedure and entanglement exhibits plateau-like 
features. {These latter reflect non-Gaussianity 
of the RT process, and are the counterpart of 
the plateaus of the single-qubit coherence in the strong 
coupling regime $g \gg 1$ \cite{galperin2006PRL}, observed in the experiment of Ref. \cite{nakamura2002PRL}.} 

Plateaus occur also when $g_A \neq g_B$, provided at least one qubit is sufficiently strongly coupled, as
shown in Fig.~\ref{fig:Cecho}(b).
The effect originates from echo pulses on the 
qubit affected by the strongly coupled fluctuator. 

\begin{figure}
\begin{center}
{\includegraphics[width=0.4\textwidth]{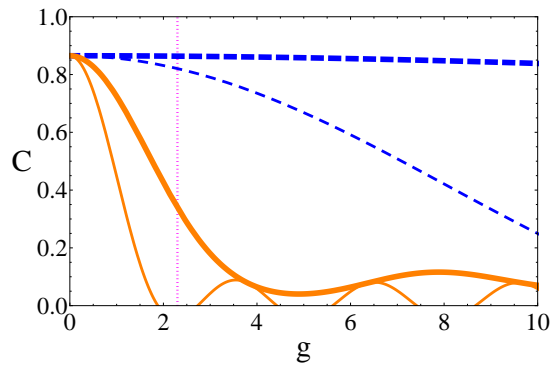}}
\caption{\label{fig:Cecho_g}\footnotesize (Color online) 
Concurrence after an echo sequence 
$C(2\Delta t)$, as a function of $g \equiv g_s$ (thick lines). 
It is shown the behavior for small pulse interval 
($\gamma \Delta t=0.1$ blue dashed line) and
for larger interval ($\gamma \Delta t=1$ 
orange solid line). 
For comparison  the concurrence in the absence of echo 
at the same time $t = 2 \Delta t$ is reported (thin dashed and
solid lines).}
\end{center}
\end{figure}

{A figure of merit for the entanglement-echo efficiency 
is the concurrence at the echo time $t=2\Delta t$.
We analyze its behavior as a function of the couplings 
$g_s \equiv g$ in Fig.~\ref{fig:Cecho_g}, and compare with 
the concurrence at the same times.}
As expected, for small pulse interval, 
$\Delta t\ll 1/\gamma$, echo 
is very effective in suppressing the noise, even 
for relatively large values of $g$. 
A richer scenario is found for $\Delta t \sim 1/\gamma$.  
In the absence of echo the concurrence at fixed 
$t= 2\Delta t$ is non-monotonic with $g$
(see Fig.~\ref{fig:Cecho_g}, thin solid line, for 
$t=1/\gamma$).   
In particular in the in the limit $g \gg 1$, 
an analytic expression can be found
the asymptotic expansion of Eq.~(\ref{singlequbitcoherence}),
$|q_\mathrm{RTN}(t)| \approx \exp{(- \gamma t/2)} \cos{(g \gamma t/2)}$. Such oscillations reflect the already discussed 
entanglement collapses and revivals as a function of time,
occurring in the regime $g>\bar{g}$. 
Oscillatory behavior of $C$ is observed also in the 
presence of the echo pulse (Fig.~\ref{fig:Cecho_g}, thick solid lines). 
Interestingly, echo preserves entanglement even when 
it vanishes in the absence of the pulse.  
However, the recovered entanglement 
might still not be sufficient for efficient realization of
quantum error correction tasks. 
Finally for increasing $\Delta t > 1/\gamma$, the echo procedure 
becomes more and more inefficiently in reducing 
detrimental effects of noise, whatsoever $g$.

\subsection{Entanglement protection by DD sequences}
The above results suggest that a sequence of pulses may preserve entanglement for longer time. Here we investigate this issue for PDD, CP and UDD. {For illustrative purposes it is 
sufficient to consider identical qubits, $g_s=g$.}

We first study how entanglement-DD performs against 
RTN in both the strong and the weak 
coupling regimes, for 
increasing pulse rate. We consider PDD sequences 
and evaluate the concurrence at fixed times $\bar t$ and 
different number $n$ of pulses equally spaced, 
$\Delta t = \bar t/n $. This quantity can be obtained from 
the exact result, Eq.(\ref{coherenceDD}), on DD of a single 
qubit from a pure dephasing RT 
fluctuator~\cite{falci2004PRA,falci2005PhysE}. Moreover, 
in order to get insight on the effect of non-Gaussianity of RTN,
we compare the above results, with their Gaussian approximation.
In this latter the coherences are obtained using 
Eq.(\ref{Gausscoherence}), where $S_s( \omega)$ is the power
spectrum of an Ornstein-Uhlenbeck process.
\begin{figure}
\begin{center}
{\includegraphics[width=0.4\textwidth]{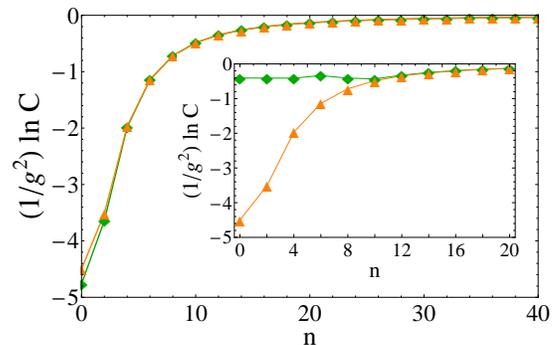}}
\caption{\label{fig:CBBGaussRTN}
\footnotesize (Color online) Comparison  $[\ln C(\bar t)]/g^2$, at the 
fixed time $\gamma \bar{t}=10$ for PDD, as a function of the number of pulses $n$ 
for RTN (green diamonds) and Gaussian (Ornstein-Uhlenbeck) 
noise (orange triangles). The values of $g$ are 0.5 (principal panel) and 5 (inset). The initial state is pure maximally entangled.}
\end{center}
\end{figure}

\begin{figure}[t!]
\begin{center}
{\includegraphics[width=0.4\textwidth]{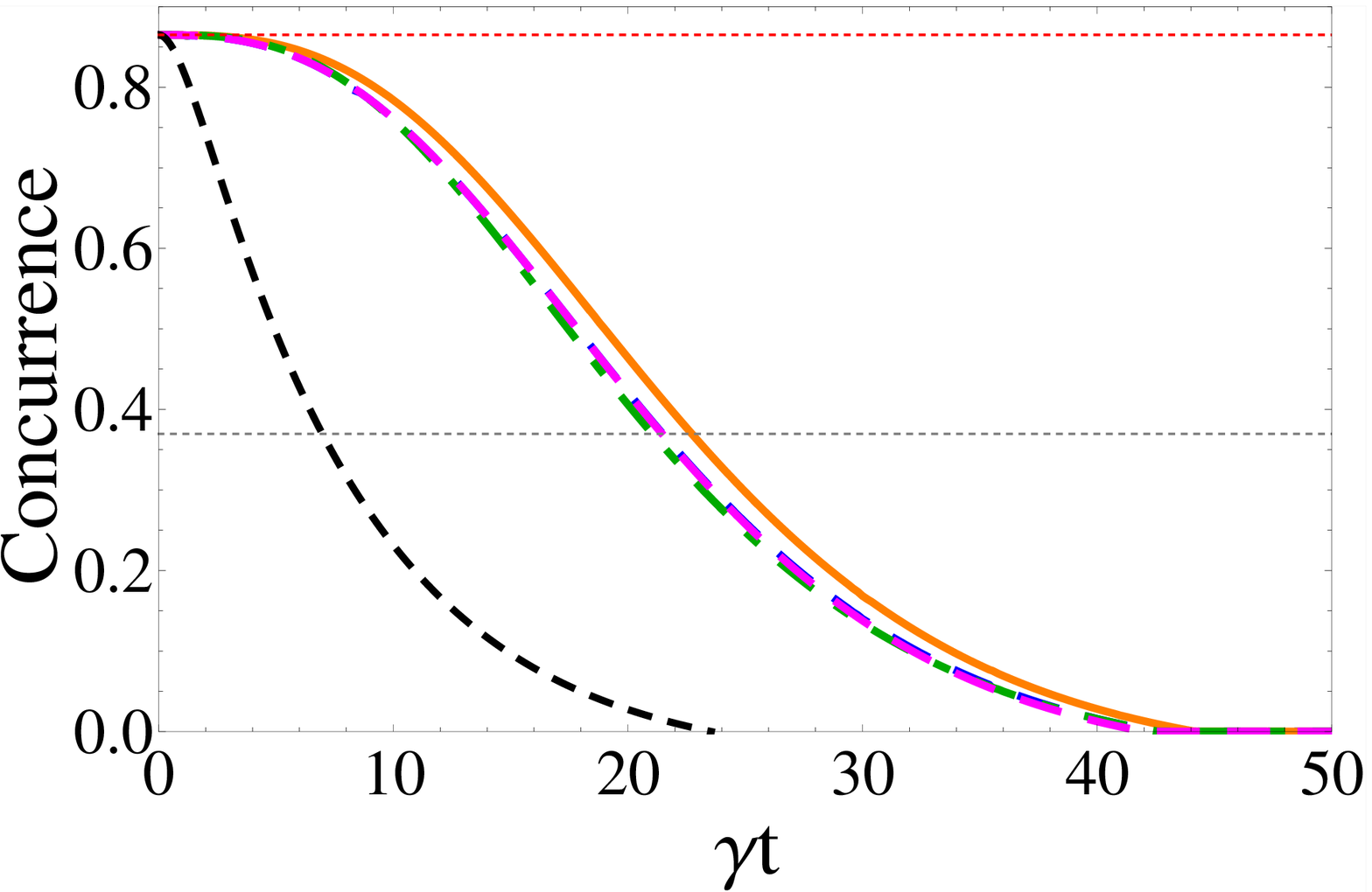}\vspace{0.5 cm}
\includegraphics[width=0.4\textwidth]{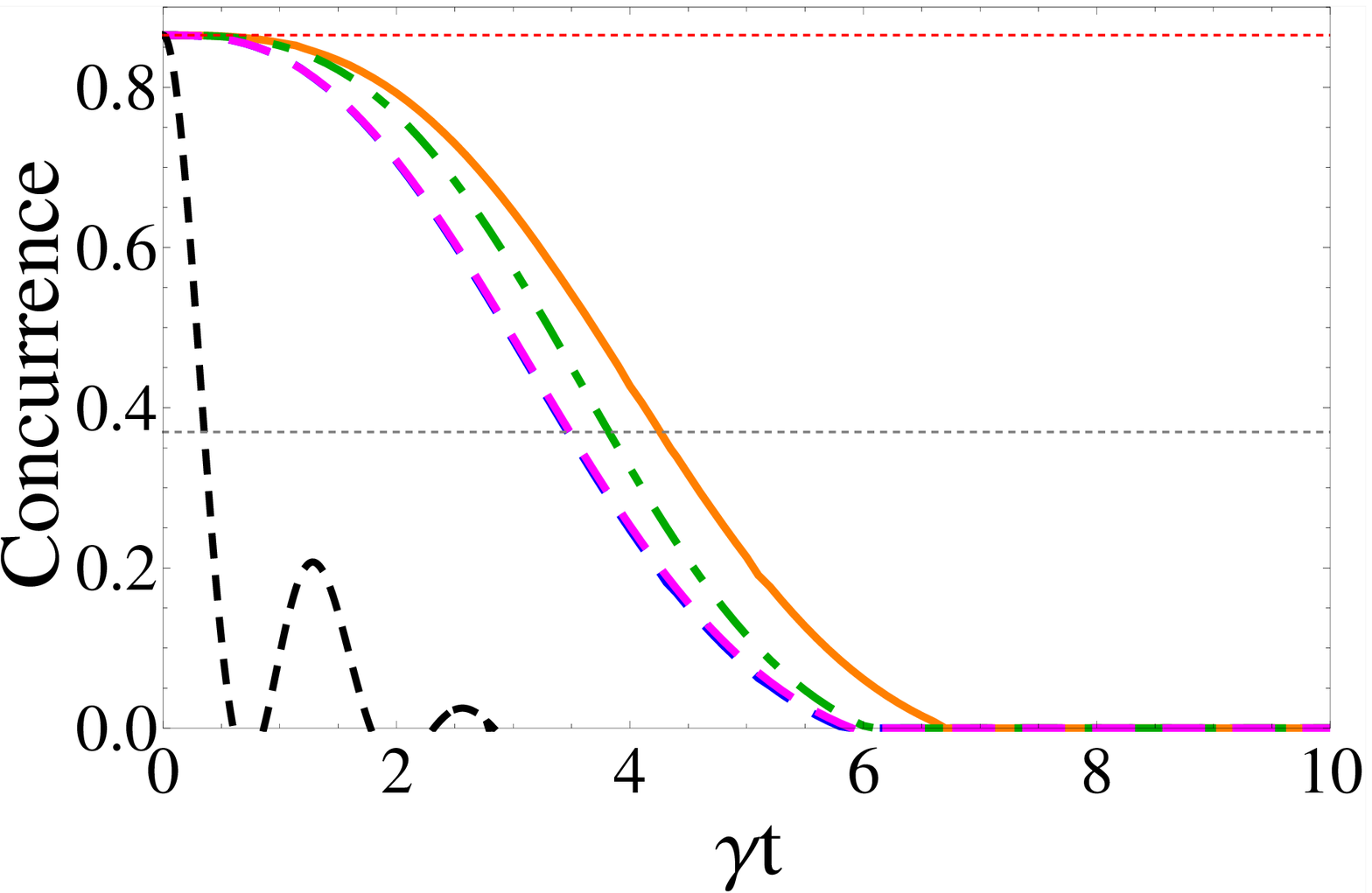}}
\caption{\label{fig:CBBvsTime}
\footnotesize (Color online) Concurrence as a function of the dimensionless
time $\gamma t$ for $g=0.5$ (top panel) and $g=5$ (bottom panel) for a fixed number of pulses $n=10$ ($\delta p_0=0$). 
Black dashed lines represent concurrences without DD sequences. 
Magenta long-dashed line, green dot-dashed line and orange solid line represent concurrences for PDD, UDD and CP, respectively. A blue long-dashed line 
is also displayed coinciding with the magenta one and representing PDD in Gaussian approximation.   
The gray dotted line at $C_\mathrm{th}\approx0.37$ represents threshold value of $C$ for Bell inequality violation. The red dotted line at the top is the initial value of concurrence.}
\end{center}
\end{figure} 
In Fig.~\ref{fig:CBBGaussRTN} we plot 
$[\ln C(\bar{t})]/g^2$ as a function of the 
(even) number of pulses $n$
for $\gamma \bar t=10$ and for an initial Bell state.
Notice that in the Gaussian approximation~\cite{viola1998PRA},
since 
$\Gamma_s(t) \propto g^2$, also the quantity 
$\ln[C(\bar{t})] = -[\Gamma_A(\bar{t})+\Gamma_B(\bar{t})] 
\propto g^2$, thereby the combination plotted in 
Fig.~\ref{fig:CBBGaussRTN} is universal. On the contrary
a dependence on $g$ related to non-Gaussianity 
appears for RTN. Actually 
for $g < 1$ (main panel) the exact result and the Gaussian 
approximation practically coincide, 
showing that the discrete nature 
of RTN is not relevant in weak coupling, 
as expected on physical grounds even  
in the absence of pulsed control~\cite{paladino2002PRL}). Instead
for $g > \bar g$ (inset) the equivalence is recovered 
only at large $n$, i.e. for sufficiently large pulse rates, 
where $\ln[C(\bar{t})]/g^2$ shows universal behavior.  
For intermediate rates $1/\Delta t \sim \gamma$, although PDD cancels dark periods and revivals 
(see Fig. \ref{fig:CnoBB}), the entanglement recovered 
is relatively small since its decay is already 
relatively slow, $\sim e^{-2\gamma t}$.

The monotonic increase of $[\ln C(\bar{t})]/g^2$ with increasing pulse rate, $1/ \Delta t$, shows how PDD suppresses 
the effect of RTN. In the universal regime 
$\gamma \Delta t \ll 1$, very frequent $\pi$-pulses 
about $\sigma_x$ coherently average out the effect of 
RTNs with $\gamma \ll 1/\Delta t$ and partially 
reconstruct the initial entanglement, independently 
of the coupling regime. This argument holds more in general
and explains why in this regime 
different statistical properties of Gaussian and non-Gaussian
processes do not produce a distinguishable effect on the 
entanglement dynamics.

The above argumentation also suggests that the equivalence at large 
pulse rates holds also for CP and UDD sequences in general. 
{This can be shown explicitly by comparing the Gaussian 
approximations with entanglement calculated using the 
numerical solution for the qubit coherences in 
the presence of RTN \cite{cywinsky2008PRB}.}
This analysis pointed out that high-order noise correlators 
in the decay factor of the qubit coherences are
suppressed by CP and UDD, the former performing 
better than the latter, 
which is by construction optimized to reduce
the second cumulant (Gaussian term). 
For the CP and UDD sequences the Gaussian approximation 
for the qubit coherences was found to be
applicable up to $g=10$ for $n=10$ pulses \cite{cywinsky2008PRB}.

Based on these considerations, in the following we study 
how entanglement is preserved as a function of time.
We consider sequences with a fixed number $n$ of pulses 
for each $t$. This would be the outcome of an experiment 
where the qubits evolve during runs of assigned duration
$t$ under the considered $n$-pulse sequence.
The overall curve is recorded from 
successive runs varying $t$, but not $n$. 
We will use the Gaussian 
approximation for the coherences under the CP and UDD sequences,
whereas for PDD we resort to the exact result form 
Eq.(\ref{coherenceDD}). 

Results, shown in Fig.~\ref{fig:CBBvsTime} for $n=10$ pulses illustrate that the DD procedures preserve entanglement at 
times longer than for its natural complete disappearance, 
independently on the coupling conditions. 
We notice that the different qualitative behaviors of the entanglement observed for $g \leq \bar g$ and $g > \bar g$
in the absence of pulses (see also Fig.(\ref{fig:CnoBB})) are canceled by DD.
Indeed, from the filter functions of the considered sequences it is easy to see that the application of $n$ pulses within
time $t$ effectively suppresses the effect of noise components below frequency $\omega_n \sim 2n/t$ \cite{cywinsky2008PRB}. 
CP slightly outperforms UDD, PDD being the less effective sequence. 
Notice how effective are pulse sequences 
in the strong coupling regime 
(lower panel of Fig.~\ref{fig:CBBvsTime}):
for instance, for $g=5$, the CP sequence keeps 
the concurrence above the
 nonlocality threshold $C_\mathrm{th}$ for times 
($\gamma t \approx 4.5$) an order 
of magnitude longer than in absence of 
pulses ($\gamma t\approx 0.4$). 
In the weak coupling regime DD extends to larger times  
the initial short times behavior. As a consequence nonlocality 
becomes more robust and the ESD times are delayed.
\begin{figure}[t!]
\begin{center}
{\includegraphics[width=0.4\textwidth]{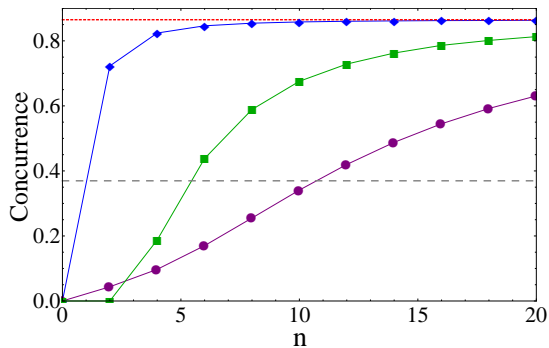}}
\caption{\label{fig:CnESD}\footnotesize (Color online) Concurrence under CP as a function of the number of pulses 
$n$ at fixed times $\gamma\bar{t}=\gamma t_\mathrm{ESD}$ for $g\leq\bar{g}$ and 
$\gamma\bar{t}=\gamma t_\mathrm{FD}$ for $g>\bar{g}$. The values of $g$ are 0.5 (purple circles),
$\bar{g}=2.3$ (blue diamonds) and 5 (green squares).
The gray dashed line at $C_\mathrm{th}\approx0.37$ represents the 
threshold value of $C$ after which there is a Bell inequality violation. The red dotted line is the initial value of concurrence.}
\end{center}
\end{figure}

Another figure of merit of the DD sequences is how many pulses are required to store the entanglement until 
the ESD-FD times. To address this issue we evaluate the concurrence under DD at the time $\bar t$ where
it definitively vanishes under free evolution.  
These times depend on the coupling conditions and are given by the ESD times for $g\leq\bar{g}$ and 
by the FD times for $g>\bar{g}$ (see Table~\ref{tabletESD}). 
This analysis is reported in Fig.~\ref{fig:CnESD} for the CP sequence, that is the most efficient pulse sequence. 
Notice that the minimum number of pulses to store entanglement is obtained for $g=\bar{g}=2.3$: already two pulses
allow exceeding the threshold $C_\mathrm{th}$, while ten pulses give $C=0.857$. i.e. an error $\sim 0.9\%$ with respect to
 the initial value $C(0)=0.865$. 
This is due to the fact that for $g=\bar{g}$ 
entanglement disappears completely in the
shortest time (see table \ref{tabletESD}). This means that 
small pulse intervals are needed, but on the other hand 
few of them very efficiently recover the large amount of 
entanglement initially lost in the absence of pulses, as 
shown by the fast initial increase of $C(\gamma \bar{t})$
in Fig.~\ref{fig:CnESD}.

\section{Dynamical decoupling of $1/f$ noise}
In this Section we investigate the efficiency of PDD, CP and 
UDD sequences to preserve entanglement between two 
superconducting qubits in the presence of pure-dephasing 
Broad Band Colored Noise (BBCN). This is the common instance in
solid-state implementations of quantum processors. We 
refer to the experimental situation reported in 
Refs. \cite{bylander2011NatPhys,yan2012PRB} where 
magnetic flux noise on a flux-type superconducting qubit 
has been well characterized. Consistent measurements of 
$1/f$-type power laws have been reported at $0.2$-$20$ MHz 
and in the range $0.01$-$100$ Hz. Based on these results, 
we consider flux noise
$S_\Phi (\omega) = A_\Phi/ (2 \pi \omega)$, 
extending between $1$~Hz and 
$10$~MHz with amplitude 
$A_\Phi=(1.7 \times 10^{-6} \, \Phi_0)^2$ 
($\Phi_0 = h/(2e)$ is the magnetic flux quantum).
For the above noise amplitude and $\gamma_i \in[1, 10^7]$~Hz, 
Eq.(\ref{eq:1overf-powerspec}) 
yields  
$\sigma \approx 2\pi\times10^7$ Hz. We 
attribute this figure of noise to a large number 
($N=10^4$) of impurities with a narrow distribution of 
couplings about the average $\bar{v}/2\pi=0.2$ MHz. 
We remark that $\gamma_M=10^7$ Hz is a soft UV-cutoff, the spectrum
decaying as $\omega^{-2}$ at larger frequencies. 
\begin{figure}
\begin{center}
{\includegraphics[width=0.4\textwidth]{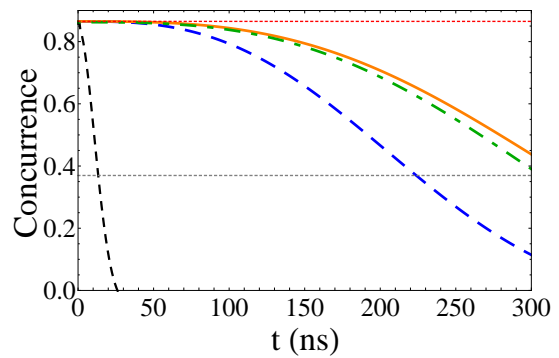}\vspace{0.5 cm}
\includegraphics[width=0.4\textwidth]{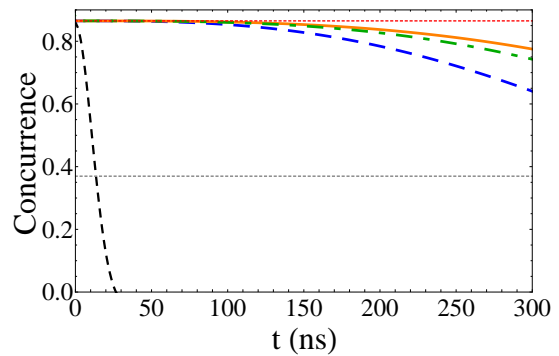}}
\caption{\label{fig:Ct1overf}\footnotesize 
(Color online) Concurrence as a function of time $t$ 
at a fixed number of pulses $n =4$ (top panel) and $n =10$ (bottom panel). 
Blue long-dashed line, green dot-dashed line and orange solid line represent concurrences for PDD, UDD and CP, respectively. 
Black dashed line is the concurrence in absence of pulses, 
entanglement disappearing at $t_\mathrm{ESD}\approx 27 
\,\mathrm{ns}$. 
The gray dotted line at $C_\mathrm{th}\approx0.37$ 
is the threshold value of $C$ for after Bell inequality 
violation. 
The top red dotted line is the initial value of concurrence.
The $1/f$ noise figures are given in the text.}
\end{center}
\end{figure}

The single qubit coherence $q_{1/f}(t)=\Pi_{i=1}^{N}q_i(t)$ is obtained as the product of $N$ coherences $q_i(t)$ associated to 
each RTN, which in absence of pulses is 
given by Eq.~(\ref{singlequbitcoherence}). 
The resulting concurrence Eq.~(\ref{Kt}) allows to 
estimate the ESD-time which will be used as a benchmark for 
entanglement recovery. Using quasi-static 
$1/f$-noise \cite{falci2005PRL,ithier2005PRB} we have 
$t_\mathrm{ESD} \approx 1/[\sigma \sqrt{\ln{(4|ab|r/(1-r))}}]$
\cite{palermocatania2010PRA}, and the corresponding figure 
for the initial EWL state and $\sigma$ 
we consider is $t_\mathrm{ESD}\approx 27\,\mathrm{ns}$. 

Results for other sequences are found following the same steps. 
For PDD we use the exact $q_\mathrm{PDD}(t)$ of Eq.~(\ref{coherenceDD}), whereas CP and UDD sequences are studied in 
the Gaussian approximation Eq. (\ref{Gausscoherence}) 
for the qubit coherences, evaluated with   
$S^{1/f}(\omega)$ Eq.(\ref{eq:1overf-powerspec})  
and the appropriate filter functions. In each case we 
finally find the concurrences from Eq.~(\ref{Kt}).

The effectiveness of the considered DD sequences is 
analyzed in Fig.~\ref{fig:Ct1overf} where 
we display the concurrence vs time 
$t \in [0,0.3 \, \mu\mathrm{s}]$ for sequences 
with fixed number of pulses ($n=4$ and $n=10$).
The CP sequence shows remarkable performances, since 
already  4 pulses allow to store entanglement for times $t \sim 500\,\mathrm{ns}$, much 
larger than $t_\mathrm{ESD}\approx 27 
\,\mathrm{ns}$. For $10$ CP pulses 
entanglement is preserved until $t \sim 900\,\mathrm{ns}$.
It is worth stressing that the concurrence exceeds the 
the non-locality threshold $C_\mathrm{th}$ for times scales 
relevant for computation, namely $t \sim 300\,\mathrm{ns}$
with $4$-pulses CP, to be compared with 
$t \approx 13\,\mathrm{ns}$ in the absence of pulsed control.
Operating $4$ CP pulses in $300\,\mathrm{ns}$ is 
absolutely within current 
technologies~\cite{bylander2011NatPhys} and there is 
room for improvement by higher pulse rates. 
We find that the CP sequence outperforms PDD and UDD,
despite of the expectation that UDD gains from its strong low-frequency filtering properties. 
Therefore, CP sequence interestingly works best for entanglement protection in the no-cutoff case.
This circumstance, which is also observed in experiments, is due in our model 
to the $\sim \omega^{-2}$ tail of the 
spectrum for frequencies larger than $\gamma_M$.

\begin{figure}
\begin{center}
{\includegraphics[width=0.4\textwidth]{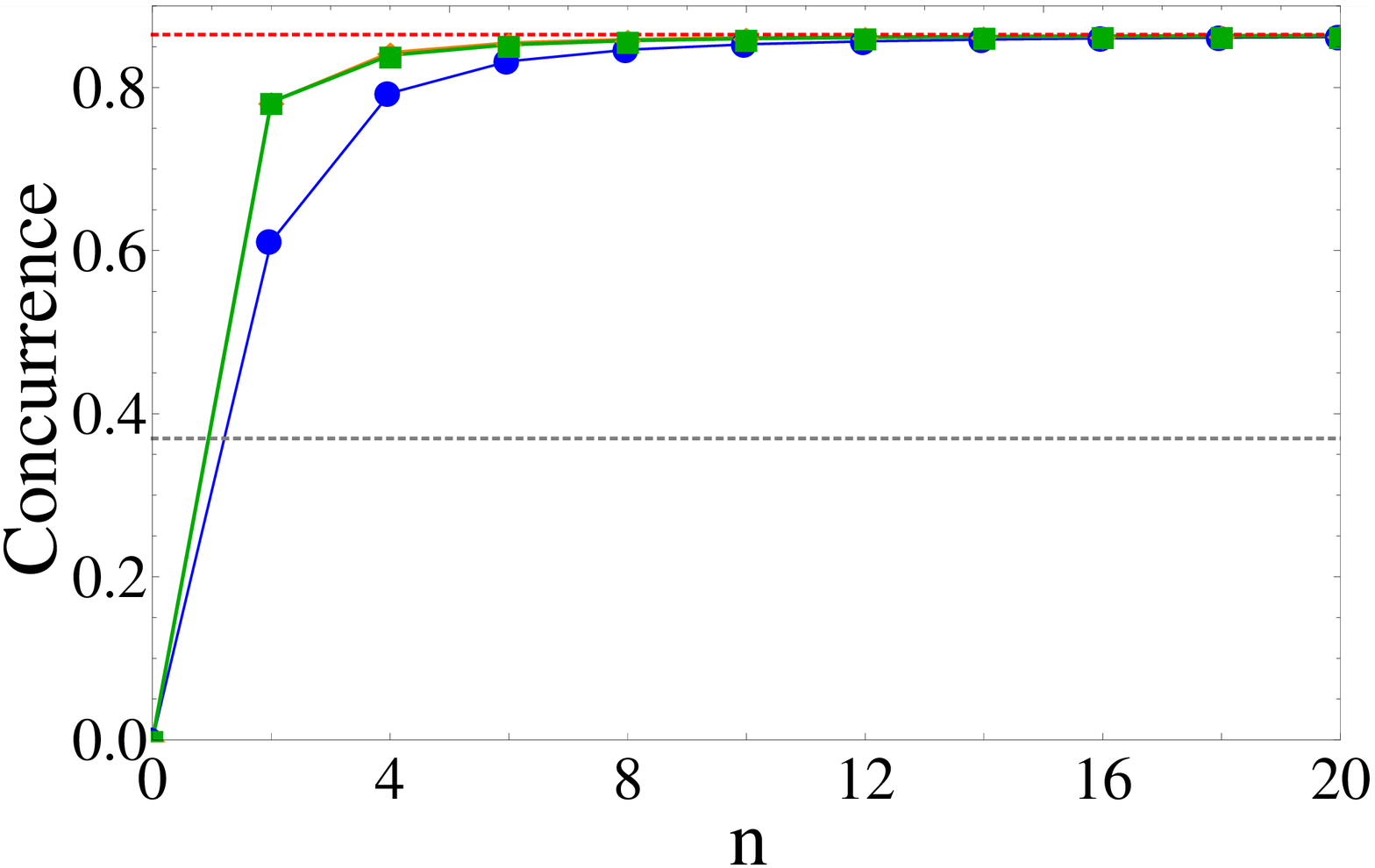}\vspace{0.5 cm}
\includegraphics[width=0.4\textwidth]{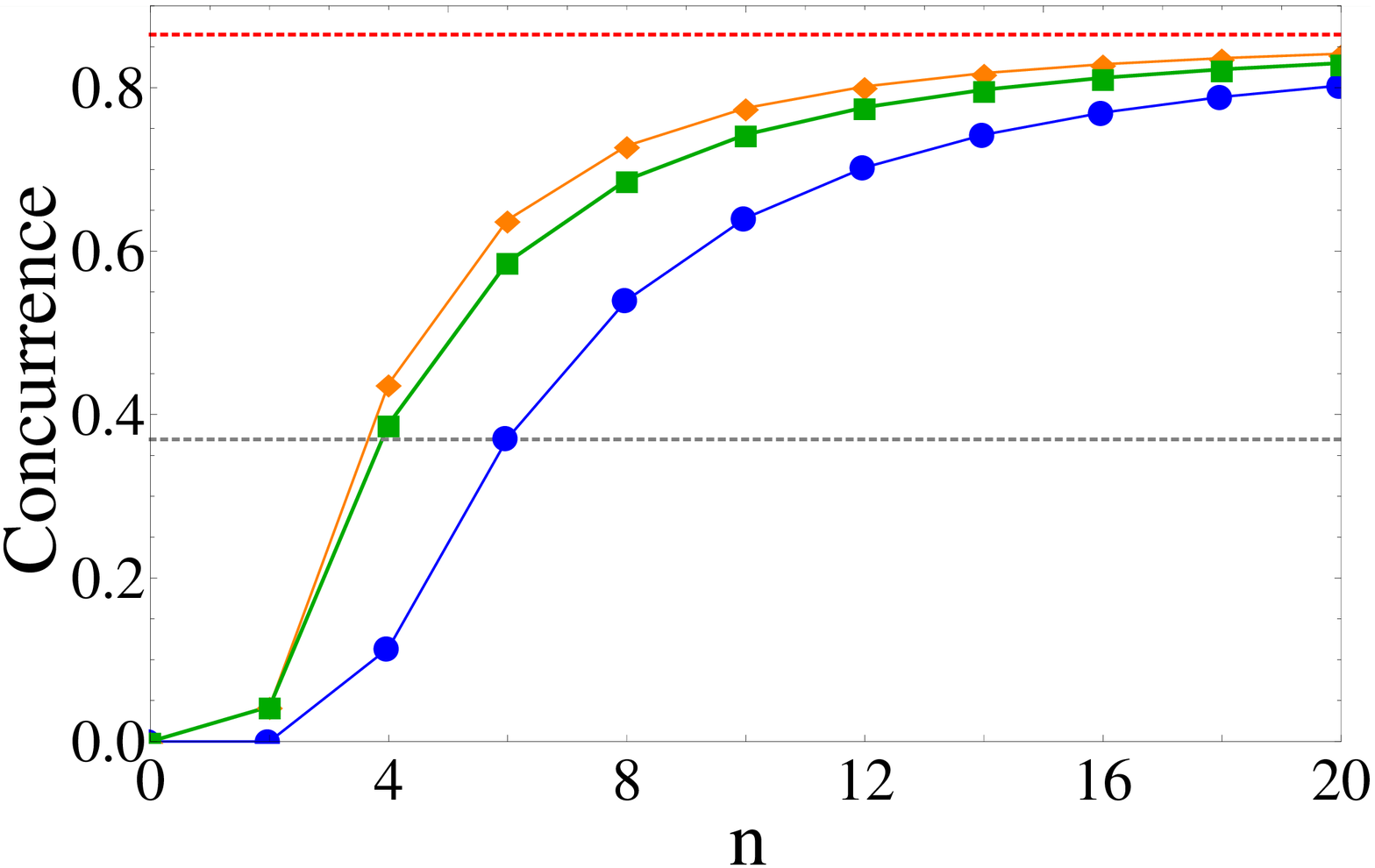}}
\caption{\label{fig:Cn1overf}\footnotesize (Color online) Concurrence as a function of the number of pulses $n$ 
at $\bar t_1=0.1 \mu$s (top panel) and $\bar t_2=0.3 \mu$s (bottom panel). Blue circles, green squares and orange diamonds represent concurrences for PDD, UDD and CP, respectively. The CP points are superposed to the UDD ones.
The gray dashed line at $C_\mathrm{th}\approx0.37$ represents the threshold value of $C$ after which there is a Bell inequality violation. 
The red dotted line at the top is the initial value of concurrence.}
\end{center}
\end{figure}

In Fig.~\ref{fig:Cn1overf} we study how increasing the 
number of pulses a high degree of 
entanglement and nonlocality at given times 
($\bar t_1=0.1\,\mu \,\mathrm{s}$ and 
$\bar t_2=0.3\,\mu \,\mathrm{s}$), both much larger than
$t_\mathrm{ESD}$. Notice that we have short $t_{ESD}$ since  
the problem addressed in this paper, namely $1/f$ noise at pure dephasing, is the ``worse scenario'' for dephasing.
In a single qubit effects of 
noise with sharp UV-cutoff $\gamma_M$ 
are strongly suppressed if $2 n/\bar t> \gamma_M$, 
UDD yielding high-fidelity at short times and CP 
being more efficient at longer times \cite{cywinsky2008PRB,pasini2010PRA,liu2013PRA}.
We find qualitatively the same behavior for entanglement, 
except that we consider a soft-UV cutoff obtaining 
only a partial noise suppression.
Notice that at $\bar{t}= 0.1\,\mu$s a concurrence larger 
than $C_\mathrm{th}$ is recovered 
with only $n=2$ pulses by both CP and UDD sequences. They 
yield $C \sim 0.8$, the corresponding 
PDD yielding slightly worse results, being 
a two-pulse echo with not so large $\Delta t=0.5/\gamma_M$. 
The advantage of using the CP sequence emerges at longer times. 
At $0.3$ $\mu$s entanglement is kept above the nonlocality threshold $C_\mathrm{th}$ with only $4$ pulses, 
Fig.~\ref{fig:Cn1overf} (bottom panel). For this protocol 
the interval between pulses, 
$\Delta t_1=\Delta t_4 =37.5\,$ns and 
$\Delta t_2=\Delta t_3 =75\,$ns, are well within 
current experimental resources. 
On the contrary PDD with $n=4$, with
$\Delta t= 4/(3 \gamma_M)$, has a worse performance
since the effect of fluctuators switching at $\gamma \sim \gamma_M$ is not cancelled. 
Notice finally that for the parameter we consider
strongly coupled fluctuators, such that $\gamma \ll \bar v = 0.4 \pi$MHz, are effectively averaged out since we consider
higher pulse rates, $1/\Delta t \gg \bar v$~\cite{falci2004PRA}. 

It should be noted that for noises with strong low-frequency component, like $1/f$ noise,
there is a strong inhomogeneous broadening effect which is efficiently removed by echo and 
by CP sequence, in the multi-pulse case. As visible in the results shown, PDD indeed shows up 
as an inconvenient sequence compared to a CP sequence with the same number of pulses.

{\em DD sequence efficiency.}--
The performance of each DD sequence in preserving entanglement until time $t$ can be quantified by evaluating
the efficiency defined as 
\begin{equation}\label{efficiencyDD}
\mathcal{E}_{n}(t)=C_n^{\mathrm{DD}}(t)/C(0),
\end{equation}
where $C_n^{\mathrm{DD}}(t)$ is the concurrence for a given DD sequence with $n$ pulses at time $t$. 
Another figure of merit is the fidelity to the initial EWL state given by \cite{nielsenchuang}
\begin{equation}\label{fidelity}
\mathcal{F}_{n}(t)=\mathrm{Tr}\sqrt{\sqrt{\rho_i}\rho_n^{\mathrm{DD}}(t)\sqrt{\rho_i}},
\end{equation}
where $\rho_i$ is the initial EWL Eq. (\ref{EWLstates}), 
while $\rho_n^{\mathrm{DD}}(t)$ is the evolved state 
under DD control.

We list in table~\ref{tableperformance} the values 
of efficiency $\mathcal{E}_{n}(t)$ and fidelity $\mathcal{F}_n(t)$ for the CP sequence in the
presence of $1/f$ noise at times 
$\bar t_1= 0.1 \,\mu$s, $\bar t_2=0.3 \, \mu$s and 
different $n$.
\begin{table}
\begin{center}
\footnotesize
\begin{tabular}{c|c c}\hline\hline
$n$ & $\bar t_1=0.1 \mu$s & $\bar t_2=0.3 \mu$s \\
\hline
2 & (0.905,\ 0.99451) & (0.049,\ 0.84100)  \\
4 & (0.975,\ 0.99945) & (0.506,\ 0.93661)  \\
6 & (0.989,\ 0.99988) & (0.734,\ 0.97428)  \\
8 & (0.994,\ 0.99996) & (0.843,\ 0.98796)  \\
10 & (0.996,\ 0.99998) & (0.896,\ 0.99370)  \\
\hline\hline
\end{tabular}
\caption{\label{tableperformance}\footnotesize Values of CP sequence efficiency $\mathcal{E}_{n}(t)$ and fidelity $\mathcal{F}_n(t)$, displayed as $(\mathcal{E},\mathcal{F})$,
at the indicated times and different number of pulses $n$, 
for initial EWL states with $r=r_\mathrm{exp}=0.91$ and $|a|=1/\sqrt{2}$ and $1/f$ noise figures discussed in the
text.}
\end{center}
\end{table}
We observe that the CP sequence can be very effective in protecting entanglement against $1/f$ 
noise up to times of the order of the total 
typical duration of two-qubit gate 
sequences\cite{schoelkopf2009Nature}.
Moreover these results provide evidence that 
the CP sequence is the most effective, 
the entanglement being kept over the threshold $C_\mathrm{th}$ 
of Bell inequality violation up to times 
allowing for instance, for secure quantum cryptography.

\section{Conclusions}
In this paper we investigated  DD techniques protecting 
entanglement and nonlocality in a system of two noninteracting 
qubits subject to strong low-frequency noise, 
with $1/f$ spectrum.
We focused on a realistic setup of distributed quantum memory,
operating DD by different sequences (PDD, CP, UDD) of pulses. 
We used typical noise and control figures from  
superconducting qubits where these sequences have been 
recently used to mitigate single-qubit dephasing due 
to low-frequency 
noise~\cite{bylander2011NatPhys,gustavsson2012PRL}.
Differently from other proposals, where DD of entanglement 
between two qubits is realized by nonlocal control~\cite{muhktar2010PRA1} or nested~\cite{muhktar2010PRA2,wang2011PRA} sequences, 
here we considered simple DD pulse sequences acting 
locally on each qubit.

In particular we focused on the question whether it is possible 
to preserve non-locality for a class of mixed states,  
allowing for applications to Quantum Technologies. To this 
end, first of all, we demonstrated  
a closed relation - Eq. (\ref{BversusC}) - between an entanglement quantifier (concurrence) and nonlocality, quantified 
by the Bell function. This 
dynamical connection 
among different quantum correlation quantifiers, which is 
the main mathematical 
result of our work, 
is valid during the system evolution 
under any local pure-dephasing qubit-environment interaction
of systems prepared in EWL states. 
It implies that a threshold value of concurrence 
$C_\mathrm{th}$ exists, which depends on the purity of the 
initial state, above which the two-qubit system exhibits 
nonlocal correlations with certainty, providing a benchmark 
for the performance of strategies of entanglement protection.  
Entanglement can be preserved by local pulses since the two 
qubits are independent and affected by local 
pure dephasing noise, an effect recently pointed out in Refs~\cite{darrigo2012arxiv,darrigo2013physscripta,darrigo2014}.

We first studied pulsed control in the presence of a single 
RTN identifying a variety of behaviors depending
on the qubit-fluctuator coupling $g$.
We found that effects of two-pulse echo depend on whether 
$g$ is smaller or larger than a  threshold value $\bar{g}$,  
marking the existence of entanglement revivals 
in the absence of pulses.
When $g\leq\bar{g}$ the ESD time is delayed, whereas 
when $g>\bar{g}$ the dynamical features of revivals and dark 
periods are canceled out by the echoes, 
entanglement exhibiting plateau-like behaviors. 
Multi pulse sequences turn out to be more efficient in 
entanglement preservation, prolonging the ESD time and 
allowing to keep $C > C_\mathrm{th}$. 

We then studied entanglement DD from $1/f$ noise with 
amplitude typically observed in superconducting nanocircuits. 
We have shown that entanglement 
and its nonlocal features can be stored very efficiently up to 
times an order of magnitude longer than natural entanglement disappearance times, which is the physical message of our work. 
These storage times are long enough to perform two-qubit 
quantum operations with pulse timings of current experimental reach \cite{bylander2011NatPhys,gustavsson2012PRL}.  

In this manuscript we considered hard-pulses, 
i.e. ideal instantaneous pulses.
The real finite-pulse duration could be included in this analysis by appropriately modifying the filter functions, 
as in Ref.\cite{bylander2011NatPhys} where authors assumed square pulses (about $7$~ns duration). We do not expect that this analysis would give substantial modification of our results, also in consideration of the fact that we concentrated on rather small number of pulses \cite{cywinsky2008PRB}.
Instead, we expect that numerical optimization of control 
pulses \cite{Mottonen2006,Gordon2008,Rebentrost2009} 
and realistic bounded amplitude control~\cite{Gorman2012} 
may further improve the efficiency of the investigated 
entanglement memory element. Room for improvement is also 
expected since, as we find in our simulations 
for both RTN 
and pure dephasing $1/f$ noise, and as observed in
experiments in the latter case, the CP sequence 
outperforms PDD and UDD.
Finally, it would be  interesting to extend our work to study 
decoupling from correlated low-frequency noise sources 
acting on both qubits, 
which produce distinctive decoherence effects 
in realistic solid-state quantum 
hardware~\cite{ka:208-darrigo-njp-corrnoise}.

\begin{acknowledgments}
R.L.F. and A.D. acknowledge support from the Centro Siciliano di Fisica Nucleare e Struttura della Materia (Catania). 
R.L.F. acknowledges support from the Brazilian funding agency CAPES [Pesquisador Visitante Especial-Grant No. 108/2012].
R.L.F. also acknowledges discussions with P. Caldara. This work is partially supported 
by MIUR through Grant No. PON02-00355-3391233,
Tecnologie per l'ENERGia e l'Efficienza energETICa - ENERGETIC.
\end{acknowledgments}

\appendix
\section{Qubit coherence in the presence of RTN and under PDD}
In this appendix we report the analytic form of the single-qubit coherence 
in the presence of pure dephasing random telegraph noise under PDD with an even number of pulses $q_\mathrm{PDD}(t)$
derived in \textcite{falci2004PRA,falci2005PhysE}. 
Here we put $\alpha=\sqrt{1-g^2}$ and remind that pulses are applied at times $t=n \Delta t$
\begin{equation}\label{coherenceDD}
q_\mathrm{PDD}(t)=\frac{E^{\frac{n}{2}}e^{-\frac{\gamma t}{2}}}{|\alpha|^{n}}\left\{\frac{I_-}{EG_2}[E_1+\delta p_0(E_3+iE_2)]+I_+\right\} \, ,
\end{equation}
where
\begin{eqnarray}\label{functionsforF1}
E_0&=&|\alpha|^2\left|\cosh\left(\frac{\alpha \gamma t}{2n}\right)\right|^2+(1+g^2)\left|\sinh\left(\frac{\alpha \gamma t}{2n}\right)\right|^2,\nonumber\\
E_1&=&2\Re\left\{\alpha\cosh\left(\frac{\alpha \gamma t}{2n}\right)\sinh\left(\frac{\alpha \gamma t}{2n}\right)^\ast\right\},\nonumber\\
E_2&=&-2g\left|\sinh\left(\frac{\alpha \gamma t}{2n}\right)\right|^2,\nonumber\\
E_3&=&-2g\Im\left\{\alpha\cosh\left(\frac{\alpha \gamma t}{2n}\right)\sinh\left(\frac{\alpha \gamma t}{2n}\right)^\ast\right\},\nonumber\\
E&=&\left[|E_0|^2-\sum_{i=1}^3 |E_i|^2\right]^{1/2}, \nonumber \\
G_1&=&\frac{E_0}{E},\quad G_2=\frac{\left[\sum_{i=1}^3 |E_i|^2\right]^{1/2}}{E} ,\nonumber\\
I_\pm&=&[(G_1+G_2)^{n/2}\pm(G_1-G_2)^{n/2}]/2.\nonumber
\end{eqnarray}
The symbols $\Re$ and $\Im$ indicate, respectively, the real part and the imaginary part of the complex number.


\begin{thebibliography}{85}
\expandafter\ifx\csname natexlab\endcsname\relax\def\natexlab#1{#1}\fi
\expandafter\ifx\csname bibnamefont\endcsname\relax
  \def\bibnamefont#1{#1}\fi
\expandafter\ifx\csname bibfnamefont\endcsname\relax
  \def\bibfnamefont#1{#1}\fi
\expandafter\ifx\csname citenamefont\endcsname\relax
  \def\citenamefont#1{#1}\fi
\expandafter\ifx\csname url\endcsname\relax
  \def\url#1{\texttt{#1}}\fi
\expandafter\ifx\csname urlprefix\endcsname\relax\def\urlprefix{URL }\fi
\providecommand{\bibinfo}[2]{#2}
\providecommand{\eprint}[2][]{\url{#2}}

\bibitem[{\citenamefont{Zurek}(2003)}]{kr:203-zurek-rmp-decoherence}
\bibinfo{author}{\bibfnamefont{W.~H.} \bibnamefont{Zurek}},
  \bibinfo{journal}{Rev. Mod. Phys.} \textbf{\bibinfo{volume}{75}},
  \bibinfo{pages}{715} (\bibinfo{year}{2003}).

\bibitem[{\citenamefont{Yu and Eberly}(2009)}]{yu2009Science}
\bibinfo{author}{\bibfnamefont{T.}~\bibnamefont{Yu}} \bibnamefont{and}
  \bibinfo{author}{\bibfnamefont{J.~H.} \bibnamefont{Eberly}},
  \bibinfo{journal}{Science} \textbf{\bibinfo{volume}{323}},
  \bibinfo{pages}{598} (\bibinfo{year}{2009}).

\bibitem[{\citenamefont{Paladino et~al.}(2014)\citenamefont{Paladino, Galperin,
  Falci, and Altshuler}}]{PaladinoReview2014}
\bibinfo{author}{\bibfnamefont{E.}~\bibnamefont{Paladino}},
  \bibinfo{author}{\bibfnamefont{Y.}~\bibnamefont{Galperin}},
  \bibinfo{author}{\bibfnamefont{G.}~\bibnamefont{Falci}}, \bibnamefont{and}
  \bibinfo{author}{\bibfnamefont{B.}~\bibnamefont{Altshuler}},
  \bibinfo{journal}{Rev. Mod. Phys.} \textbf{\bibinfo{volume}{86}},
  \bibinfo{pages}{361} (\bibinfo{year}{2014}).

\bibitem[{\citenamefont{Xiang et~al.}(2013)\citenamefont{Xiang, Ashhab, You,
  and Nori}}]{norireview}
\bibinfo{author}{\bibfnamefont{Z.-L.} \bibnamefont{Xiang}},
  \bibinfo{author}{\bibfnamefont{S.}~\bibnamefont{Ashhab}},
  \bibinfo{author}{\bibfnamefont{J.}~\bibnamefont{You}}, \bibnamefont{and}
  \bibinfo{author}{\bibfnamefont{F.}~\bibnamefont{Nori}},
  \bibinfo{journal}{Rev. Mod. Phys.} \textbf{\bibinfo{volume}{85}},
  \bibinfo{pages}{623} (\bibinfo{year}{2013}).

\bibitem[{\citenamefont{Ladd et~al.}(2010)\citenamefont{Ladd, Jelezko,
  Laflamme, Nakamura, Monroe, and {O'Brien}}}]{obrienreview}
\bibinfo{author}{\bibfnamefont{T.~D.} \bibnamefont{Ladd}},
  \bibinfo{author}{\bibfnamefont{F.}~\bibnamefont{Jelezko}},
  \bibinfo{author}{\bibfnamefont{R.}~\bibnamefont{Laflamme}},
  \bibinfo{author}{\bibfnamefont{Y.}~\bibnamefont{Nakamura}},
  \bibinfo{author}{\bibfnamefont{C.}~\bibnamefont{Monroe}}, \bibnamefont{and}
  \bibinfo{author}{\bibfnamefont{J.~L.} \bibnamefont{{O'Brien}}},
  \bibinfo{journal}{Nature} \textbf{\bibinfo{volume}{464}}, \bibinfo{pages}{45}
  (\bibinfo{year}{2010}).

\bibitem[{\citenamefont{Acin et~al.}(2006)\citenamefont{Acin, Gisin, and
  Masanes}}]{acin2006PRL}
\bibinfo{author}{\bibfnamefont{A.}~\bibnamefont{Acin}},
  \bibinfo{author}{\bibfnamefont{N.}~\bibnamefont{Gisin}}, \bibnamefont{and}
  \bibinfo{author}{\bibfnamefont{L.}~\bibnamefont{Masanes}},
  \bibinfo{journal}{Phys. Rev. Lett.} \textbf{\bibinfo{volume}{97}},
  \bibinfo{pages}{120405} (\bibinfo{year}{2006}).

\bibitem[{\citenamefont{Gisin and Thew}(2007)}]{gisin2007natphoton}
\bibinfo{author}{\bibfnamefont{N.}~\bibnamefont{Gisin}} \bibnamefont{and}
  \bibinfo{author}{\bibfnamefont{R.}~\bibnamefont{Thew}},
  \bibinfo{journal}{Nature Photon.} \textbf{\bibinfo{volume}{1}},
  \bibinfo{pages}{165} (\bibinfo{year}{2007}).

\bibitem[{\citenamefont{Pironio et~al.}(2010)\citenamefont{Pironio, Ac{\'{i}}n,
  Massar, de~la Giroday, Matsukevich, Maunz, Olmschenk, Hayes, Luo, Manning
  et~al.}}]{pironio2010Nature}
\bibinfo{author}{\bibfnamefont{S.}~\bibnamefont{Pironio}},
  \bibinfo{author}{\bibfnamefont{A.}~\bibnamefont{Ac{\'{i}}n}},
  \bibinfo{author}{\bibfnamefont{S.}~\bibnamefont{Massar}},
  \bibinfo{author}{\bibfnamefont{A.~B.} \bibnamefont{de~la Giroday}},
  \bibinfo{author}{\bibfnamefont{D.~N.} \bibnamefont{Matsukevich}},
  \bibinfo{author}{\bibfnamefont{P.}~\bibnamefont{Maunz}},
  \bibinfo{author}{\bibfnamefont{S.}~\bibnamefont{Olmschenk}},
  \bibinfo{author}{\bibfnamefont{D.}~\bibnamefont{Hayes}},
  \bibinfo{author}{\bibfnamefont{L.}~\bibnamefont{Luo}},
  \bibinfo{author}{\bibfnamefont{T.~A.} \bibnamefont{Manning}},
  \bibnamefont{and} \bibinfo{author}{\bibfnamefont{C.}~\bibnamefont{Monroe}},
  \bibinfo{journal}{Nature}
  \textbf{\bibinfo{volume}{464}}, \bibinfo{pages}{1021} (\bibinfo{year}{2010}).

\bibitem[{\citenamefont{Lucamarini et~al.}(2012)\citenamefont{Lucamarini,
  Vallone, Gianani, Giuseppe, and Mataloni}}]{lucamarini2012PRA}
\bibinfo{author}{\bibfnamefont{M.}~\bibnamefont{Lucamarini}},
  \bibinfo{author}{\bibfnamefont{G.}~\bibnamefont{Vallone}},
  \bibinfo{author}{\bibfnamefont{I.}~\bibnamefont{Gianani}},
  \bibinfo{author}{\bibfnamefont{G.~D.} \bibnamefont{Giuseppe}},
  \bibnamefont{and} \bibinfo{author}{\bibfnamefont{P.}~\bibnamefont{Mataloni}},
  \bibinfo{journal}{Phys. Rev. A} \textbf{\bibinfo{volume}{86}},
  \bibinfo{pages}{032325} (\bibinfo{year}{2012}).

\bibitem[{\citenamefont{Horodecki et~al.}(2009)\citenamefont{Horodecki,
  Horodecki, Horodecki, and Horodecki}}]{horodecki2009RMP}
\bibinfo{author}{\bibfnamefont{R.}~\bibnamefont{Horodecki}},
  \bibinfo{author}{\bibfnamefont{P.}~\bibnamefont{Horodecki}},
  \bibinfo{author}{\bibfnamefont{M.}~\bibnamefont{Horodecki}},
  \bibnamefont{and}
  \bibinfo{author}{\bibfnamefont{K.}~\bibnamefont{Horodecki}},
  \bibinfo{journal}{Rev. Mod. Phys.} \textbf{\bibinfo{volume}{81}},
  \bibinfo{pages}{865} (\bibinfo{year}{2009}).

\bibitem[{\citenamefont{Bellomo et~al.}(2008)\citenamefont{Bellomo, {Lo
  Franco}, and Compagno}}]{bellomo2008bell}
\bibinfo{author}{\bibfnamefont{B.}~\bibnamefont{Bellomo}},
  \bibinfo{author}{\bibfnamefont{R.}~\bibnamefont{{Lo Franco}}},
  \bibnamefont{and} \bibinfo{author}{\bibfnamefont{G.}~\bibnamefont{Compagno}},
  \bibinfo{journal}{Phys. Rev. A} \textbf{\bibinfo{volume}{78}},
  \bibinfo{pages}{062309} (\bibinfo{year}{2008}).

\bibitem[{\citenamefont{Viola and Lloyd}(1998)}]{viola1998PRA}
\bibinfo{author}{\bibfnamefont{L.}~\bibnamefont{Viola}} \bibnamefont{and}
  \bibinfo{author}{\bibfnamefont{S.}~\bibnamefont{Lloyd}},
  \bibinfo{journal}{Phys. Rev. A} \textbf{\bibinfo{volume}{58}},
  \bibinfo{pages}{2733} (\bibinfo{year}{1998}).

\bibitem[{\citenamefont{Vandersypen and Chuang}(2005)}]{vandersypen2005RMP}
\bibinfo{author}{\bibfnamefont{L.~M.~K.} \bibnamefont{Vandersypen}}
  \bibnamefont{and} \bibinfo{author}{\bibfnamefont{I.~L.}
  \bibnamefont{Chuang}}, \bibinfo{journal}{Rev. Mod. Phys.}
  \textbf{\bibinfo{volume}{76}}, \bibinfo{pages}{1037} (\bibinfo{year}{2005}).

\bibitem[{\citenamefont{Wiseman and Milburn}(2010)}]{wiseman2010book}
\bibinfo{author}{\bibfnamefont{H.~M.} \bibnamefont{Wiseman}} \bibnamefont{and}
  \bibinfo{author}{\bibfnamefont{G.~J.} \bibnamefont{Milburn}},
  \emph{\bibinfo{title}{Quantum Measurement and Control}}
  (\bibinfo{publisher}{Cambridge University Press, USA}, \bibinfo{address}{New
  York}, \bibinfo{year}{2010}).

\bibitem[{\citenamefont{Mims}(1972)}]{Mims1972}
\bibinfo{author}{\bibfnamefont{W.~B.} \bibnamefont{Mims}},
  \emph{\bibinfo{title}{Electron Paramagnetic Resonance}}
  (\bibinfo{publisher}{New York: Plenum}, \bibinfo{year}{1972}), pp.
  \bibinfo{pages}{263--351}.

\bibitem[{\citenamefont{Viola et~al.}(1999)\citenamefont{Viola, Knill, and
  Lloyd}}]{viola1999PRL}
\bibinfo{author}{\bibfnamefont{L.}~\bibnamefont{Viola}},
  \bibinfo{author}{\bibfnamefont{E.}~\bibnamefont{Knill}}, \bibnamefont{and}
  \bibinfo{author}{\bibfnamefont{S.}~\bibnamefont{Lloyd}},
  \bibinfo{journal}{Phys. Rev. Lett.} \textbf{\bibinfo{volume}{82}},
  \bibinfo{pages}{2417} (\bibinfo{year}{1999}).

\bibitem[{\citenamefont{Carr and Purcell}(1954)}]{carr1954PhysRev}
\bibinfo{author}{\bibfnamefont{H.~Y.} \bibnamefont{Carr}} \bibnamefont{and}
  \bibinfo{author}{\bibfnamefont{E.~M.} \bibnamefont{Purcell}},
  \bibinfo{journal}{Phys. Rev.} \textbf{\bibinfo{volume}{94}},
  \bibinfo{pages}{630} (\bibinfo{year}{1954}).

\bibitem[{\citenamefont{Uhrig}(2007)}]{uhrig2007PRL}
\bibinfo{author}{\bibfnamefont{G.~S.} \bibnamefont{Uhrig}},
  \bibinfo{journal}{Phys. Rev. Lett.} \textbf{\bibinfo{volume}{98}},
  \bibinfo{pages}{100504} (\bibinfo{year}{2007}).

\bibitem[{\citenamefont{Yang and R.-B-Liu}(2008)}]{yang2008PRL}
\bibinfo{author}{\bibfnamefont{W.}~\bibnamefont{Yang}} \bibnamefont{and}
  \bibinfo{author}{\bibnamefont{R.-B-Liu}}, \bibinfo{journal}{Phys. Rev. Lett.}
  \textbf{\bibinfo{volume}{101}}, \bibinfo{pages}{180403}
  (\bibinfo{year}{2008}).

\bibitem[{\citenamefont{Biercuk et~al.}(2011)\citenamefont{Biercuk, Doherty,
  and Uys}}]{biercuk2011JPB}
\bibinfo{author}{\bibfnamefont{M.~J.} \bibnamefont{Biercuk}},
  \bibinfo{author}{\bibfnamefont{A.~C.} \bibnamefont{Doherty}},
  \bibnamefont{and} \bibinfo{author}{\bibfnamefont{H.}~\bibnamefont{Uys}},
  \bibinfo{journal}{J. Phys. B: At. Mol. Opt. Phys.}
  \textbf{\bibinfo{volume}{44}}, \bibinfo{pages}{154002}
  (\bibinfo{year}{2011}).

\bibitem[{\citenamefont{Falci et~al.}(2004)\citenamefont{Falci, D'Arrigo,
  Mastellone, and Paladino}}]{falci2004PRA}
\bibinfo{author}{\bibfnamefont{G.}~\bibnamefont{Falci}},
  \bibinfo{author}{\bibfnamefont{A.}~\bibnamefont{D'Arrigo}},
  \bibinfo{author}{\bibfnamefont{A.}~\bibnamefont{Mastellone}},
  \bibnamefont{and} \bibinfo{author}{\bibfnamefont{E.}~\bibnamefont{Paladino}},
  \bibinfo{journal}{Phys. Rev. A} \textbf{\bibinfo{volume}{70}},
  \bibinfo{pages}{040101(R)} (\bibinfo{year}{2004}).

\bibitem[{\citenamefont{Rebentrost et~al.}(2009)\citenamefont{Rebentrost,
  Serban, Schulte-Herbrüggen, and Wilhelm}}]{Rebentrost2009}
\bibinfo{author}{\bibfnamefont{P.}~\bibnamefont{Rebentrost}},
  \bibinfo{author}{\bibfnamefont{I.}~\bibnamefont{Serban}},
  \bibinfo{author}{\bibfnamefont{T.}~\bibnamefont{Schulte-Herbrüggen}},
  \bibnamefont{and} \bibinfo{author}{\bibfnamefont{F.~K.}
  \bibnamefont{Wilhelm}}, \bibinfo{journal}{Phys. Rev. Lett.}
  \textbf{\bibinfo{volume}{102}}, \bibinfo{pages}{090401}
  (\bibinfo{year}{2009}).

\bibitem[{\citenamefont{Lutchyn et~al.}(2008)\citenamefont{Lutchyn,
  Cywi{\'{n}}ski, Nave, and {Das Sarma}}}]{lutchyn2008PRB}
\bibinfo{author}{\bibfnamefont{R.~M.} \bibnamefont{Lutchyn}},
  \bibinfo{author}{\bibfnamefont{L.}~\bibnamefont{Cywi{\'{n}}ski}},
  \bibinfo{author}{\bibfnamefont{C.~P.} \bibnamefont{Nave}}, \bibnamefont{and}
  \bibinfo{author}{\bibfnamefont{S.}~\bibnamefont{{Das Sarma}}},
  \bibinfo{journal}{Phys. Rev. B} \textbf{\bibinfo{volume}{78}},
  \bibinfo{pages}{024508} (\bibinfo{year}{2008}).

\bibitem[{\citenamefont{Faoro and Viola}(2004)}]{faoro2004PRL}
\bibinfo{author}{\bibfnamefont{L.}~\bibnamefont{Faoro}} \bibnamefont{and}
  \bibinfo{author}{\bibfnamefont{L.}~\bibnamefont{Viola}},
  \bibinfo{journal}{Phys. Rev. Lett.} \textbf{\bibinfo{volume}{92}},
  \bibinfo{pages}{117905} (\bibinfo{year}{2004}).

\bibitem[{\citenamefont{Bergli and Faoro}(2007)}]{bergli2007PRB}
\bibinfo{author}{\bibfnamefont{J.}~\bibnamefont{Bergli}} \bibnamefont{and}
  \bibinfo{author}{\bibfnamefont{L.}~\bibnamefont{Faoro}},
  \bibinfo{journal}{Phys. Rev. B} \textbf{\bibinfo{volume}{75}},
  \bibinfo{pages}{054515} (\bibinfo{year}{2007}).

\bibitem[{\citenamefont{Cheng et~al.}(2008)\citenamefont{Cheng, Wang, and
  Joynt}}]{Cheng2008}
\bibinfo{author}{\bibfnamefont{B.}~\bibnamefont{Cheng}},
  \bibinfo{author}{\bibfnamefont{Q.~H.} \bibnamefont{Wang}}, \bibnamefont{and}
  \bibinfo{author}{\bibfnamefont{R.}~\bibnamefont{Joynt}},
  \bibinfo{journal}{Phys. Rev. A} \textbf{\bibinfo{volume}{78}},
  \bibinfo{pages}{022313} (\bibinfo{year}{2008}).

\bibitem[{\citenamefont{Gutmann et~al.}(2005)\citenamefont{Gutmann, Wilhelm,
  Kaminsky, and Lloyd}}]{Gutmann2005}
\bibinfo{author}{\bibfnamefont{H.}~\bibnamefont{Gutmann}},
  \bibinfo{author}{\bibfnamefont{F.~K.} \bibnamefont{Wilhelm}},
  \bibinfo{author}{\bibfnamefont{W.~M.} \bibnamefont{Kaminsky}},
  \bibnamefont{and} \bibinfo{author}{\bibfnamefont{S.}~\bibnamefont{Lloyd}},
  \bibinfo{journal}{Phys. Rev. A} \textbf{\bibinfo{volume}{71}},
  \bibinfo{pages}{020302(R)} (\bibinfo{year}{2005}).

\bibitem[{\citenamefont{Cywi{\'{n}}ski
  et~al.}(2008)\citenamefont{Cywi{\'{n}}ski, Lutchyn, Nave, and {Das
  Sarma}}}]{cywinsky2008PRB}
\bibinfo{author}{\bibfnamefont{L.}~\bibnamefont{Cywi{\'{n}}ski}},
  \bibinfo{author}{\bibfnamefont{R.~M.} \bibnamefont{Lutchyn}},
  \bibinfo{author}{\bibfnamefont{C.~P.} \bibnamefont{Nave}}, \bibnamefont{and}
  \bibinfo{author}{\bibfnamefont{S.}~\bibnamefont{{Das Sarma}}},
  \bibinfo{journal}{Phys. Rev. B} \textbf{\bibinfo{volume}{77}},
  \bibinfo{pages}{174509} (\bibinfo{year}{2008}).

\bibitem[{\citenamefont{Du et~al.}(2009)\citenamefont{Du, Rong, Zhao, Wang,
  Yang, and Liu}}]{du2009nature}
\bibinfo{author}{\bibfnamefont{J.~F.} \bibnamefont{Du}},
  \bibinfo{author}{\bibfnamefont{X.}~\bibnamefont{Rong}},
  \bibinfo{author}{\bibfnamefont{N.}~\bibnamefont{Zhao}},
  \bibinfo{author}{\bibfnamefont{Y.}~\bibnamefont{Wang}},
  \bibinfo{author}{\bibfnamefont{J.}~\bibnamefont{Yang}}, \bibnamefont{and}
  \bibinfo{author}{\bibfnamefont{R.~B.} \bibnamefont{Liu}},
  \bibinfo{journal}{Nature} \textbf{\bibinfo{volume}{461}},
  \bibinfo{pages}{1265} (\bibinfo{year}{2009}).

\bibitem[{\citenamefont{Lee et~al.}(2008)\citenamefont{Lee, Witzel, and
  Das~Sarma}}]{Lee2008}
\bibinfo{author}{\bibfnamefont{B.}~\bibnamefont{Lee}},
  \bibinfo{author}{\bibfnamefont{W.~M.} \bibnamefont{Witzel}},
  \bibnamefont{and}
  \bibinfo{author}{\bibfnamefont{S.}~\bibnamefont{Das~Sarma}},
  \bibinfo{journal}{Phys. Rev. Lett.} \textbf{\bibinfo{volume}{100}},
  \bibinfo{pages}{160505} (\bibinfo{year}{2008}).

\bibitem[{\citenamefont{Nakamura et~al.}(2002)\citenamefont{Nakamura, Pashkin,
  Yamamoto, and Tsai}}]{nakamura2002PRL}
\bibinfo{author}{\bibfnamefont{Y.}~\bibnamefont{Nakamura}},
  \bibinfo{author}{\bibfnamefont{Y.~A.} \bibnamefont{Pashkin}},
  \bibinfo{author}{\bibfnamefont{T.}~\bibnamefont{Yamamoto}}, \bibnamefont{and}
  \bibinfo{author}{\bibfnamefont{J.~S.} \bibnamefont{Tsai}},
  \bibinfo{journal}{Phys. Rev. Lett.} \textbf{\bibinfo{volume}{88}},
  \bibinfo{pages}{047901} (\bibinfo{year}{2002}).

\bibitem[{\citenamefont{Bertet et~al.}(2005)\citenamefont{Bertet, Chiorescu,
  Burkard, Semba, Harmans, DiVincenzo, and Mooij}}]{Bertet2005}
\bibinfo{author}{\bibfnamefont{P.}~\bibnamefont{Bertet}},
  \bibinfo{author}{\bibfnamefont{I.}~\bibnamefont{Chiorescu}},
  \bibinfo{author}{\bibfnamefont{G.}~\bibnamefont{Burkard}},
  \bibinfo{author}{\bibfnamefont{K.}~\bibnamefont{Semba}},
  \bibinfo{author}{\bibfnamefont{C.~J. P.~M.} \bibnamefont{Harmans}},
  \bibinfo{author}{\bibfnamefont{D.}~\bibnamefont{DiVincenzo}},
  \bibnamefont{and} \bibinfo{author}{\bibfnamefont{J.~E.} \bibnamefont{Mooij}},
  \bibinfo{journal}{Phys. Rev. Lett.} \textbf{\bibinfo{volume}{95}},
  \bibinfo{pages}{257002} (\bibinfo{year}{2005}).

\bibitem[{\citenamefont{Ithier et~al.}(2005)\citenamefont{Ithier, Collin,
  Joyez, Meeson, Vion, Esteve, Chiarello, Shnirman, Makhlin, Schrief
  et~al.}}]{ithier2005PRB}
\bibinfo{author}{\bibfnamefont{G.}~\bibnamefont{Ithier}},
  \bibinfo{author}{\bibfnamefont{E.}~\bibnamefont{Collin}},
  \bibinfo{author}{\bibfnamefont{P.}~\bibnamefont{Joyez}},
  \bibinfo{author}{\bibfnamefont{P.~J.} \bibnamefont{Meeson}},
  \bibinfo{author}{\bibfnamefont{D.}~\bibnamefont{Vion}},
  \bibinfo{author}{\bibfnamefont{D.}~\bibnamefont{Esteve}},
  \bibinfo{author}{\bibfnamefont{F.}~\bibnamefont{Chiarello}},
  \bibinfo{author}{\bibfnamefont{A.}~\bibnamefont{Shnirman}},
  \bibinfo{author}{\bibfnamefont{Y.}~\bibnamefont{Makhlin}},
  \bibinfo{author}{\bibfnamefont{J.}~\bibnamefont{Schrief}},
\bibnamefont{and} \bibinfo{author}{\bibfnamefont{G.} \bibnamefont{Sch{\"{o}}n}},
  \bibinfo{journal}{Phys. Rev. B}
  \textbf{\bibinfo{volume}{72}}, \bibinfo{pages}{134519}
  (\bibinfo{year}{2005}).

\bibitem[{\citenamefont{Yoshihara et~al.}(2006)\citenamefont{Yoshihara,
  Harrabi, Niskanen, Nakamura, and Tsai}}]{Yoshihara2006}
\bibinfo{author}{\bibfnamefont{F.}~\bibnamefont{Yoshihara}},
  \bibinfo{author}{\bibfnamefont{K.}~\bibnamefont{Harrabi}},
  \bibinfo{author}{\bibfnamefont{A.~O.} \bibnamefont{Niskanen}},
  \bibinfo{author}{\bibfnamefont{Y.}~\bibnamefont{Nakamura}}, \bibnamefont{and}
  \bibinfo{author}{\bibfnamefont{J.~S.} \bibnamefont{Tsai}},
  \bibinfo{journal}{Phys. Rev. Lett.} \textbf{\bibinfo{volume}{97}},
  \bibinfo{pages}{167001} (\bibinfo{year}{2006}).

\bibitem[{\citenamefont{Bylander et~al.}(2011)}]{bylander2011NatPhys}
\bibinfo{author}{\bibfnamefont{J.}~\bibnamefont{Bylander}}
  \bibnamefont{et~al.}, \bibinfo{journal}{Nature Phys.}
  \textbf{\bibinfo{volume}{7}}, \bibinfo{pages}{565} (\bibinfo{year}{2011}).

\bibitem[{\citenamefont{Yan et~al.}(2012)\citenamefont{Yan, Bylander,
  Gustavsson, Yoshihara, Harrabi, Cory, Orlando, Nakamura, Tsai, and
  Oliver}}]{yan2012PRB}
\bibinfo{author}{\bibfnamefont{F.}~\bibnamefont{Yan}},
  \bibinfo{author}{\bibfnamefont{J.}~\bibnamefont{Bylander}},
  \bibinfo{author}{\bibfnamefont{S.}~\bibnamefont{Gustavsson}},
  \bibinfo{author}{\bibfnamefont{F.}~\bibnamefont{Yoshihara}},
  \bibinfo{author}{\bibfnamefont{K.}~\bibnamefont{Harrabi}},
  \bibinfo{author}{\bibfnamefont{D.~G.} \bibnamefont{Cory}},
  \bibinfo{author}{\bibfnamefont{T.~P.} \bibnamefont{Orlando}},
  \bibinfo{author}{\bibfnamefont{Y.}~\bibnamefont{Nakamura}},
  \bibinfo{author}{\bibfnamefont{J.-S.} \bibnamefont{Tsai}}, \bibnamefont{and}
  \bibinfo{author}{\bibfnamefont{W.~D.} \bibnamefont{Oliver}},
  \bibinfo{journal}{Phys. Rev. B} \textbf{\bibinfo{volume}{85}},
  \bibinfo{pages}{174521} (\bibinfo{year}{2012}).

\bibitem[{\citenamefont{Yuge et~al.}(2011)\citenamefont{Yuge, Sasaki, and
  Hirayama}}]{Yuge2011PRL}
\bibinfo{author}{\bibfnamefont{T.}~\bibnamefont{Yuge}},
  \bibinfo{author}{\bibfnamefont{S.}~\bibnamefont{Sasaki}}, \bibnamefont{and}
  \bibinfo{author}{\bibfnamefont{Y.}~\bibnamefont{Hirayama}},
  \bibinfo{journal}{Phys. Rev. Lett.} \textbf{\bibinfo{volume}{107}},
  \bibinfo{pages}{170504} (\bibinfo{year}{2011}).

\bibitem[{\citenamefont{Mukhtar
  et~al.}(2010{\natexlab{a}})\citenamefont{Mukhtar, Saw, Soh, and
  Gong}}]{muhktar2010PRA1}
\bibinfo{author}{\bibfnamefont{M.}~\bibnamefont{Mukhtar}},
  \bibinfo{author}{\bibfnamefont{T.~B.} \bibnamefont{Saw}},
  \bibinfo{author}{\bibfnamefont{W.~T.} \bibnamefont{Soh}}, \bibnamefont{and}
  \bibinfo{author}{\bibfnamefont{J.}~\bibnamefont{Gong}},
  \bibinfo{journal}{Phys. Rev. A} \textbf{\bibinfo{volume}{81}},
  \bibinfo{pages}{012331} (\bibinfo{year}{2010}{\natexlab{a}}).

\bibitem[{\citenamefont{Mukhtar
  et~al.}(2010{\natexlab{b}})\citenamefont{Mukhtar, Soh, Saw, and
  Gong}}]{muhktar2010PRA2}
\bibinfo{author}{\bibfnamefont{M.}~\bibnamefont{Mukhtar}},
  \bibinfo{author}{\bibfnamefont{W.~T.} \bibnamefont{Soh}},
  \bibinfo{author}{\bibfnamefont{T.~B.} \bibnamefont{Saw}}, \bibnamefont{and}
  \bibinfo{author}{\bibfnamefont{J.}~\bibnamefont{Gong}},
  \bibinfo{journal}{Phys. Rev. A} \textbf{\bibinfo{volume}{82}},
  \bibinfo{pages}{052338} (\bibinfo{year}{2010}{\natexlab{b}}).

\bibitem[{\citenamefont{Wang and Liu}(2011)}]{wang2011PRA}
\bibinfo{author}{\bibfnamefont{Z.-Y.} \bibnamefont{Wang}} \bibnamefont{and}
  \bibinfo{author}{\bibfnamefont{R.-B.} \bibnamefont{Liu}},
  \bibinfo{journal}{Phys. Rev. A} \textbf{\bibinfo{volume}{83}},
  \bibinfo{pages}{022306} (\bibinfo{year}{2011}).

\bibitem[{\citenamefont{Pand et~al.}(2011)\citenamefont{Pand, Z.-R-Xi, and
  Gong}}]{pan2011JPB}
\bibinfo{author}{\bibfnamefont{Y.}~\bibnamefont{Pand}},
  \bibinfo{author}{\bibnamefont{Z.-R-Xi}}, \bibnamefont{and}
  \bibinfo{author}{\bibfnamefont{J.}~\bibnamefont{Gong}}, \bibinfo{journal}{J.
  Phys. B: At. Mol. Opt. Phys.} \textbf{\bibinfo{volume}{44}},
  \bibinfo{pages}{175501} (\bibinfo{year}{2011}).

\bibitem[{\citenamefont{Gustavsson et~al.}(2012)\citenamefont{Gustavsson, Yan,
  Bylander, Yoshihara, Nakamura, Orlando, and Oliver}}]{gustavsson2012PRL}
\bibinfo{author}{\bibfnamefont{S.}~\bibnamefont{Gustavsson}},
  \bibinfo{author}{\bibfnamefont{F.}~\bibnamefont{Yan}},
  \bibinfo{author}{\bibfnamefont{J.}~\bibnamefont{Bylander}},
  \bibinfo{author}{\bibfnamefont{F.}~\bibnamefont{Yoshihara}},
  \bibinfo{author}{\bibfnamefont{Y.}~\bibnamefont{Nakamura}},
  \bibinfo{author}{\bibfnamefont{T.~P.} \bibnamefont{Orlando}},
  \bibnamefont{and} \bibinfo{author}{\bibfnamefont{W.~D.}
  \bibnamefont{Oliver}}, \bibinfo{journal}{Phys. Rev. Lett.}
  \textbf{\bibinfo{volume}{109}}, \bibinfo{pages}{010502}
  (\bibinfo{year}{2012}).

\bibitem[{\citenamefont{Wang et~al.}(2011)\citenamefont{Wang, Rong, Feng, Xu,
  Chong, Su, Gong, and Du}}]{wang2007PRL}
\bibinfo{author}{\bibfnamefont{Y.}~\bibnamefont{Wang}},
  \bibinfo{author}{\bibfnamefont{X.}~\bibnamefont{Rong}},
  \bibinfo{author}{\bibfnamefont{P.}~\bibnamefont{Feng}},
  \bibinfo{author}{\bibfnamefont{W.}~\bibnamefont{Xu}},
  \bibinfo{author}{\bibfnamefont{B.}~\bibnamefont{Chong}},
  \bibinfo{author}{\bibfnamefont{J.-H.} \bibnamefont{Su}},
  \bibinfo{author}{\bibfnamefont{J.}~\bibnamefont{Gong}}, \bibnamefont{and}
  \bibinfo{author}{\bibfnamefont{J.}~\bibnamefont{Du}}, \bibinfo{journal}{Phys.
  Rev. Lett.} \textbf{\bibinfo{volume}{106}}, \bibinfo{pages}{040501}
  (\bibinfo{year}{2011}).

\bibitem[{\citenamefont{Roy et~al.}(2011)\citenamefont{Roy, Mahesh, and
  Agarwal}}]{roy2011PRA}
\bibinfo{author}{\bibfnamefont{S.~S.} \bibnamefont{Roy}},
  \bibinfo{author}{\bibfnamefont{T.~S.} \bibnamefont{Mahesh}},
  \bibnamefont{and} \bibinfo{author}{\bibfnamefont{G.~S.}
  \bibnamefont{Agarwal}}, \bibinfo{journal}{Phys. Rev. A}
  \textbf{\bibinfo{volume}{83}}, \bibinfo{pages}{062326}
  (\bibinfo{year}{2011}).

\bibitem[{\citenamefont{Shulman et~al.}(2012)\citenamefont{Shulman, Dial,
  Harvey, Bluhm, Umansky, and Yacoby}}]{shulman2012Science}
\bibinfo{author}{\bibfnamefont{M.~D.} \bibnamefont{Shulman}},
  \bibinfo{author}{\bibfnamefont{O.~E.} \bibnamefont{Dial}},
  \bibinfo{author}{\bibfnamefont{S.~P.} \bibnamefont{Harvey}},
  \bibinfo{author}{\bibfnamefont{H.}~\bibnamefont{Bluhm}},
  \bibinfo{author}{\bibfnamefont{V.}~\bibnamefont{Umansky}}, \bibnamefont{and}
  \bibinfo{author}{\bibfnamefont{A.}~\bibnamefont{Yacoby}},
  \bibinfo{journal}{Science} \textbf{\bibinfo{volume}{336}},
  \bibinfo{pages}{202} (\bibinfo{year}{2012}).

\bibitem[{\citenamefont{Dolde et~al.}(2013)\citenamefont{Dolde, Jakobi,
  Naydenov, Zhao, Pezzagna, Trautmann, Meijer, Neumann, Jelezko, and
  Wrachtrup}}]{dolde2013NatPhys}
\bibinfo{author}{\bibfnamefont{F.}~\bibnamefont{Dolde}},
  \bibinfo{author}{\bibfnamefont{I.}~\bibnamefont{Jakobi}},
  \bibinfo{author}{\bibfnamefont{B.}~\bibnamefont{Naydenov}},
  \bibinfo{author}{\bibfnamefont{N.}~\bibnamefont{Zhao}},
  \bibinfo{author}{\bibfnamefont{S.}~\bibnamefont{Pezzagna}},
  \bibinfo{author}{\bibfnamefont{C.}~\bibnamefont{Trautmann}},
  \bibinfo{author}{\bibfnamefont{J.}~\bibnamefont{Meijer}},
  \bibinfo{author}{\bibfnamefont{P.}~\bibnamefont{Neumann}},
  \bibinfo{author}{\bibfnamefont{F.}~\bibnamefont{Jelezko}}, \bibnamefont{and}
  \bibinfo{author}{\bibfnamefont{J.}~\bibnamefont{Wrachtrup}},
  \bibinfo{journal}{Nature Phys.} \textbf{\bibinfo{volume}{9}},
  \bibinfo{pages}{139} (\bibinfo{year}{2013}).

\bibitem[{\citenamefont{Wootters}(1998)}]{wootters1998PRL}
\bibinfo{author}{\bibfnamefont{W.~K.} \bibnamefont{Wootters}},
  \bibinfo{journal}{Phys. Rev. Lett.} \textbf{\bibinfo{volume}{80}},
  \bibinfo{pages}{2245} (\bibinfo{year}{1998}).

\bibitem[{\citenamefont{Horodecki et~al.}(1995)\citenamefont{Horodecki,
  Horodecki, and Horodecki}}]{horodecki1995PLA}
\bibinfo{author}{\bibfnamefont{M.}~\bibnamefont{Horodecki}},
  \bibinfo{author}{\bibfnamefont{P.}~\bibnamefont{Horodecki}},
  \bibnamefont{and}
  \bibinfo{author}{\bibfnamefont{R.}~\bibnamefont{Horodecki}},
  \bibinfo{journal}{Phys. Lett. A} \textbf{\bibinfo{volume}{200}},
  \bibinfo{pages}{340} (\bibinfo{year}{1995}).

\bibitem[{\citenamefont{Mazzola et~al.}(2010)\citenamefont{Mazzola, Bellomo,
  {Lo Franco}, and Compagno}}]{mazzolapalermo2010PRA}
\bibinfo{author}{\bibfnamefont{L.}~\bibnamefont{Mazzola}},
  \bibinfo{author}{\bibfnamefont{B.}~\bibnamefont{Bellomo}},
  \bibinfo{author}{\bibfnamefont{R.}~\bibnamefont{{Lo Franco}}},
  \bibnamefont{and} \bibinfo{author}{\bibfnamefont{G.}~\bibnamefont{Compagno}},
  \bibinfo{journal}{Phys. Rev. A} \textbf{\bibinfo{volume}{81}},
  \bibinfo{pages}{052116} (\bibinfo{year}{2010}).

\bibitem[{\citenamefont{Horst et~al.}(2013)\citenamefont{Horst, Bartkiewicz,
  and Miranowicz}}]{horst2013PRA}
\bibinfo{author}{\bibfnamefont{B.}~\bibnamefont{Horst}},
  \bibinfo{author}{\bibfnamefont{K.}~\bibnamefont{Bartkiewicz}},
  \bibnamefont{and}
  \bibinfo{author}{\bibfnamefont{A.}~\bibnamefont{Miranowicz}},
  \bibinfo{journal}{Phys. Rev. A} \textbf{\bibinfo{volume}{87}},
  \bibinfo{pages}{042108} (\bibinfo{year}{2013}).

\bibitem[{\citenamefont{Bartkiewicz et~al.}(2013)\citenamefont{Bartkiewicz,
  Horst, Lemr, and Miranowicz}}]{bartkiewicz2013PRA}
\bibinfo{author}{\bibfnamefont{K.}~\bibnamefont{Bartkiewicz}},
  \bibinfo{author}{\bibfnamefont{B.}~\bibnamefont{Horst}},
  \bibinfo{author}{\bibfnamefont{K.}~\bibnamefont{Lemr}}, \bibnamefont{and}
  \bibinfo{author}{\bibfnamefont{A.}~\bibnamefont{Miranowicz}},
  \bibinfo{journal}{Phys. Rev. A} \textbf{\bibinfo{volume}{88}},
  \bibinfo{pages}{052105} (\bibinfo{year}{2013}).

\bibitem[{\citenamefont{Verstraete and Wolf}(2002)}]{verstraete2002PRL}
\bibinfo{author}{\bibfnamefont{F.}~\bibnamefont{Verstraete}} \bibnamefont{and}
  \bibinfo{author}{\bibfnamefont{M.~M.} \bibnamefont{Wolf}},
  \bibinfo{journal}{Phys. Rev. Lett.} \textbf{\bibinfo{volume}{89}},
  \bibinfo{pages}{170401} (\bibinfo{year}{2002}).

\bibitem[{\citenamefont{Bellomo et~al.}(2007)\citenamefont{Bellomo, {Lo
  Franco}, and Compagno}}]{bellomo2007PRL}
\bibinfo{author}{\bibfnamefont{B.}~\bibnamefont{Bellomo}},
  \bibinfo{author}{\bibfnamefont{R.}~\bibnamefont{{Lo Franco}}},
  \bibnamefont{and} \bibinfo{author}{\bibfnamefont{G.}~\bibnamefont{Compagno}},
  \bibinfo{journal}{Phys. Rev. Lett.} \textbf{\bibinfo{volume}{99}},
  \bibinfo{pages}{160502} (\bibinfo{year}{2007}).

\bibitem[{\citenamefont{Yu and Eberly}(2007)}]{yu2007QIC}
\bibinfo{author}{\bibfnamefont{T.}~\bibnamefont{Yu}} \bibnamefont{and}
  \bibinfo{author}{\bibfnamefont{J.~H.} \bibnamefont{Eberly}},
  \bibinfo{journal}{Quantum Information and Computation}
  \textbf{\bibinfo{volume}{7}}, \bibinfo{pages}{459} (\bibinfo{year}{2007}).

\bibitem[{\citenamefont{Bellomo
  et~al.}(2010{\natexlab{a}})\citenamefont{Bellomo, {Lo Franco}, and
  Compagno}}]{bellomo2010PLA}
\bibinfo{author}{\bibfnamefont{B.}~\bibnamefont{Bellomo}},
  \bibinfo{author}{\bibfnamefont{R.}~\bibnamefont{{Lo Franco}}},
  \bibnamefont{and} \bibinfo{author}{\bibfnamefont{G.}~\bibnamefont{Compagno}},
  \bibinfo{journal}{Phys. Lett. A} \textbf{\bibinfo{volume}{374}},
  \bibinfo{pages}{3007} (\bibinfo{year}{2010}{\natexlab{a}}).

\bibitem[{\citenamefont{Miranowicz}(2004)}]{miran2004PLA}
\bibinfo{author}{\bibfnamefont{A.}~\bibnamefont{Miranowicz}},
  \bibinfo{journal}{Phys. Lett. A} \textbf{\bibinfo{volume}{327}},
  \bibinfo{pages}{272} (\bibinfo{year}{2004}).

\bibitem[{\citenamefont{Bellomo et~al.}(2009)\citenamefont{Bellomo, {Lo
  Franco}, and Compagno}}]{bellomo2008nonlocal}
\bibinfo{author}{\bibfnamefont{B.}~\bibnamefont{Bellomo}},
  \bibinfo{author}{\bibfnamefont{R.}~\bibnamefont{{Lo Franco}}},
  \bibnamefont{and} \bibinfo{author}{\bibfnamefont{G.}~\bibnamefont{Compagno}},
  \bibinfo{journal}{Adv. Science Lett.} \textbf{\bibinfo{volume}{2}},
  \bibinfo{pages}{459} (\bibinfo{year}{2009}).

\bibitem[{\citenamefont{Gisin}(1991)}]{gisin1991PLA}
\bibinfo{author}{\bibfnamefont{N.}~\bibnamefont{Gisin}},
  \bibinfo{journal}{Phys. Lett. A} \textbf{\bibinfo{volume}{154}},
  \bibinfo{pages}{201} (\bibinfo{year}{1991}).

\bibitem[{\citenamefont{DiCarlo et~al.}(2009)\citenamefont{DiCarlo, Chow,
  Gambetta, Bishop, Johnson, Schuster, Majer, Blais, Frunzio, Girvin
  et~al.}}]{schoelkopf2009Nature}
\bibinfo{author}{\bibfnamefont{L.}~\bibnamefont{DiCarlo}},
  \bibinfo{author}{\bibfnamefont{J.~M.} \bibnamefont{Chow}},
  \bibinfo{author}{\bibfnamefont{J.~M.} \bibnamefont{Gambetta}},
  \bibinfo{author}{\bibfnamefont{L.~S.} \bibnamefont{Bishop}},
  \bibinfo{author}{\bibfnamefont{B.~R.} \bibnamefont{Johnson}},
  \bibinfo{author}{\bibfnamefont{D.~I.} \bibnamefont{Schuster}},
  \bibinfo{author}{\bibfnamefont{J.}~\bibnamefont{Majer}},
  \bibinfo{author}{\bibfnamefont{A.}~\bibnamefont{Blais}},
  \bibinfo{author}{\bibfnamefont{L.}~\bibnamefont{Frunzio}},
  \bibinfo{author}{\bibfnamefont{S.~M.} \bibnamefont{Girvin}}, \bibnamefont{and} 
 \bibinfo{author}{\bibfnamefont{R.~J.} \bibnamefont{Schoelkopf}}, 
 \bibinfo{journal}{Nature}
  \textbf{\bibinfo{volume}{460}}, \bibinfo{pages}{240} (\bibinfo{year}{2009}).

\bibitem[{\citenamefont{Carlo et~al.}(2010)\citenamefont{Carlo, Reed, Sun,
  Johnson, Chow, M.Gambetta, Frunzio, Girvin, Devoret, and
  Schoelkopf}}]{Dicarlo2010}
\bibinfo{author}{\bibfnamefont{L.~D.} \bibnamefont{Carlo}},
  \bibinfo{author}{\bibfnamefont{M.~D.} \bibnamefont{Reed}},
  \bibinfo{author}{\bibfnamefont{L.}~\bibnamefont{Sun}},
  \bibinfo{author}{\bibfnamefont{B.~R.} \bibnamefont{Johnson}},
  \bibinfo{author}{\bibfnamefont{J.~M.} \bibnamefont{Chow}},
  \bibinfo{author}{\bibfnamefont{J.}~\bibnamefont{M.Gambetta}},
  \bibinfo{author}{\bibfnamefont{L.}~\bibnamefont{Frunzio}},
  \bibinfo{author}{\bibfnamefont{S.~M.} \bibnamefont{Girvin}},
  \bibinfo{author}{\bibfnamefont{M.~H.} \bibnamefont{Devoret}},
  \bibnamefont{and} \bibinfo{author}{\bibfnamefont{R.~J.}
  \bibnamefont{Schoelkopf}}, \bibinfo{journal}{Nature}
  \textbf{\bibinfo{volume}{467}}, \bibinfo{pages}{574} (\bibinfo{year}{2010}).

\bibitem[{\citenamefont{Neeley et~al.}(2010)\citenamefont{Neeley, Bialczak,
  Lenander, Lucero, Mariantoni, O'Connell, Sank, Wang, Weides, Wenner
  et~al.}}]{Neeley2010}
\bibinfo{author}{\bibfnamefont{M.}~\bibnamefont{Neeley}},
  \bibinfo{author}{\bibfnamefont{R.~C.} \bibnamefont{Bialczak}},
  \bibinfo{author}{\bibfnamefont{M.}~\bibnamefont{Lenander}},
  \bibinfo{author}{\bibfnamefont{E.}~\bibnamefont{Lucero}},
  \bibinfo{author}{\bibfnamefont{M.}~\bibnamefont{Mariantoni}},
  \bibinfo{author}{\bibfnamefont{A.~D.} \bibnamefont{O'Connell}},
  \bibinfo{author}{\bibfnamefont{D.}~\bibnamefont{Sank}},
  \bibinfo{author}{\bibfnamefont{H.}~\bibnamefont{Wang}},
  \bibinfo{author}{\bibfnamefont{M.}~\bibnamefont{Weides}},
  \bibinfo{author}{\bibfnamefont{J.}~\bibnamefont{Wenner}},
  \bibinfo{author}{\bibfnamefont{Y.}~\bibnamefont{Yin}},
  \bibinfo{author}{\bibfnamefont{T.}~\bibnamefont{Yamamoto}},
  \bibinfo{author}{\bibfnamefont{A.~N.} \bibnamefont{Cleland}}, \bibnamefont{and}
  \bibinfo{author}{\bibfnamefont{J.~M.}~\bibnamefont{Martinis}},
  \bibinfo{journal}{Nature}
  \textbf{\bibinfo{volume}{467}}, \bibinfo{pages}{570} (\bibinfo{year}{2010}).

\bibitem[{\citenamefont{Lucero et~al.}(2012)\citenamefont{Lucero, Barends,
  Chen, Kelly, Mariantoni, Megrant, O'Malley, Sank, Vainsencher, Wenner
  et~al.}}]{Lucero2012}
\bibinfo{author}{\bibfnamefont{E.}~\bibnamefont{Lucero}},
  \bibinfo{author}{\bibfnamefont{R.}~\bibnamefont{Barends}},
  \bibinfo{author}{\bibfnamefont{Y.}~\bibnamefont{Chen}},
  \bibinfo{author}{\bibfnamefont{J.}~\bibnamefont{Kelly}},
  \bibinfo{author}{\bibfnamefont{M.}~\bibnamefont{Mariantoni}},
  \bibinfo{author}{\bibfnamefont{A.}~\bibnamefont{Megrant}},
  \bibinfo{author}{\bibfnamefont{P.}~\bibnamefont{O'Malley}},
  \bibinfo{author}{\bibfnamefont{D.}~\bibnamefont{Sank}},
  \bibinfo{author}{\bibfnamefont{A.}~\bibnamefont{Vainsencher}},
  \bibinfo{author}{\bibfnamefont{J.}~\bibnamefont{Wenner}},
 \bibinfo{author}{\bibfnamefont{T.}~\bibnamefont{White}},
  \bibinfo{author}{\bibfnamefont{Y.}~\bibnamefont{Yin}},
  \bibinfo{author}{\bibfnamefont{A.~N.} \bibnamefont{Cleland}}, \bibnamefont{and}
  \bibinfo{author}{\bibfnamefont{J.~M.}~\bibnamefont{Martinis}},
  \bibinfo{journal}{Nat. Phys.}
  \textbf{\bibinfo{volume}{8}}, \bibinfo{pages}{719} (\bibinfo{year}{2012}).

\bibitem[{\citenamefont{Mariantoni et~al.}(2011)\citenamefont{Mariantoni, Wang,
  Yamamoto, Neeley, Bialczak, Chen, Lenander, Lucero, O{\'C}onnell, Sank
  et~al.}}]{Mariantoni2011}
\bibinfo{author}{\bibfnamefont{M.}~\bibnamefont{Mariantoni}},
  \bibinfo{author}{\bibfnamefont{H.}~\bibnamefont{Wang}},
  \bibinfo{author}{\bibfnamefont{T.}~\bibnamefont{Yamamoto}},
  \bibinfo{author}{\bibfnamefont{M.}~\bibnamefont{Neeley}},
  \bibinfo{author}{\bibfnamefont{R.~C.} \bibnamefont{Bialczak}},
  \bibinfo{author}{\bibfnamefont{Y.}~\bibnamefont{Chen}},
  \bibinfo{author}{\bibfnamefont{M.}~\bibnamefont{Lenander}},
  \bibinfo{author}{\bibfnamefont{E.}~\bibnamefont{Lucero}},
  \bibinfo{author}{\bibfnamefont{A.~D.} \bibnamefont{O{\'C}onnell}},
  \bibinfo{author}{\bibfnamefont{D.}~\bibnamefont{Sank}},
   \bibinfo{author}{\bibfnamefont{M.}~\bibnamefont{Weides}},
  \bibinfo{author}{\bibfnamefont{J.} \bibnamefont{Wenner}},
  \bibinfo{author}{\bibfnamefont{Y.}~\bibnamefont{Yin}},
  \bibinfo{author}{\bibfnamefont{J.}~\bibnamefont{Zhao}},
  \bibinfo{author}{\bibfnamefont{A.~N.}~\bibnamefont{Korotkov}},
  \bibinfo{author}{\bibfnamefont{A.~N.} \bibnamefont{Cleland}}, \bibnamefont{and}
  \bibinfo{author}{\bibfnamefont{J.~M.}~\bibnamefont{Martinis}},
  \bibinfo{journal}{Science} \textbf{\bibinfo{volume}{334}},
  \bibinfo{pages}{61} (\bibinfo{year}{2011}).


\bibitem[{\citenamefont{Fedorov et~al.}(2012)\citenamefont{Fedorov, Steffen,
  Baur, da~Silva, and Wallraff}}]{Fedorov2012}
\bibinfo{author}{\bibfnamefont{A.}~\bibnamefont{Fedorov}},
  \bibinfo{author}{\bibfnamefont{L.}~\bibnamefont{Steffen}},
  \bibinfo{author}{\bibfnamefont{M.}~\bibnamefont{Baur}},
  \bibinfo{author}{\bibfnamefont{M.~P.} \bibnamefont{da~Silva}},
  \bibnamefont{and} \bibinfo{author}{\bibfnamefont{A.}~\bibnamefont{Wallraff}},
  \bibinfo{journal}{Nature} \textbf{\bibinfo{volume}{481}},
  \bibinfo{pages}{170} (\bibinfo{year}{2012}).

\bibitem[{\citenamefont{Reed et~al.}(2012)\citenamefont{Reed, DiCarlo, Nigg,
  Sun, Frunzio, Girvin, and Schoelkopf}}]{Reed2012}
\bibinfo{author}{\bibfnamefont{M.~D.} \bibnamefont{Reed}},
  \bibinfo{author}{\bibfnamefont{L.}~\bibnamefont{DiCarlo}},
  \bibinfo{author}{\bibfnamefont{S.~E.} \bibnamefont{Nigg}},
  \bibinfo{author}{\bibfnamefont{L.}~\bibnamefont{Sun}},
  \bibinfo{author}{\bibfnamefont{L.}~\bibnamefont{Frunzio}},
  \bibinfo{author}{\bibfnamefont{S.~M.} \bibnamefont{Girvin}},
  \bibnamefont{and} \bibinfo{author}{\bibfnamefont{R.~J.}
  \bibnamefont{Schoelkopf}}, \bibinfo{journal}{Nature}
  \textbf{\bibinfo{volume}{482}}, \bibinfo{pages}{382} (\bibinfo{year}{2012}).

\bibitem[{\citenamefont{Chow et~al.}(2012)\citenamefont{Chow, Gambetta,
  Corcoles, Merkel, Smolin, Rigetti, Poletto, Keefe, Rothwell, Rozen
  et~al.}}]{Chow2012}
\bibinfo{author}{\bibfnamefont{J.~M.} \bibnamefont{Chow}},
  \bibinfo{author}{\bibfnamefont{J.~M.} \bibnamefont{Gambetta}},
  \bibinfo{author}{\bibfnamefont{A.~D.} \bibnamefont{Corcoles}},
  \bibinfo{author}{\bibfnamefont{S.~T.} \bibnamefont{Merkel}},
  \bibinfo{author}{\bibfnamefont{J.~A.} \bibnamefont{Smolin}},
  \bibinfo{author}{\bibfnamefont{C.}~\bibnamefont{Rigetti}},
  \bibinfo{author}{\bibfnamefont{S.}~\bibnamefont{Poletto}},
  \bibinfo{author}{\bibfnamefont{G.~A.} \bibnamefont{Keefe}},
  \bibinfo{author}{\bibfnamefont{M.~B.} \bibnamefont{Rothwell}},
  \bibinfo{author}{\bibfnamefont{J.~R.} \bibnamefont{Rozen}},
   \bibinfo{author}{\bibfnamefont{M.~B.} \bibnamefont{Ketchen}}, \bibnamefont{and}
    \bibinfo{author}{\bibfnamefont{M.} \bibnamefont{Steffen}}, 
    \bibinfo{journal}{Phys. Rev. Lett.}
  \textbf{\bibinfo{volume}{109}}, \bibinfo{pages}{060501}
  (\bibinfo{year}{2012}).

\bibitem[{\citenamefont{Rigetti et~al.}(2012)\citenamefont{Rigetti, Gambetta,
  Poletto, Plourde, Chow, Corcoles, Smolin, Merkel, Rozen, Keefe
  et~al.}}]{Rigetti2012}
\bibinfo{author}{\bibfnamefont{C.}~\bibnamefont{Rigetti}},
  \bibinfo{author}{\bibfnamefont{J.~M.} \bibnamefont{Gambetta}},
  \bibinfo{author}{\bibfnamefont{S.}~\bibnamefont{Poletto}},
  \bibinfo{author}{\bibfnamefont{B.~L.~T.} \bibnamefont{Plourde}},
  \bibinfo{author}{\bibfnamefont{J.~M.} \bibnamefont{Chow}},
  \bibinfo{author}{\bibfnamefont{A.~D.} \bibnamefont{Corcoles}},
  \bibinfo{author}{\bibfnamefont{J.~A.} \bibnamefont{Smolin}},
  \bibinfo{author}{\bibfnamefont{S.~T.} \bibnamefont{Merkel}},
  \bibinfo{author}{\bibfnamefont{J.~R.} \bibnamefont{Rozen}},
  \bibinfo{author}{\bibfnamefont{G.~A.} \bibnamefont{Keefe}}, 
  \bibinfo{author}{\bibfnamefont{M.~B.} \bibnamefont{Rothwell}},
   \bibinfo{author}{\bibfnamefont{M.~B.} \bibnamefont{Ketchen}}, \bibnamefont{and}
    \bibinfo{author}{\bibfnamefont{M.} \bibnamefont{Steffen}}, 
    \bibinfo{journal}{Phys. Rev. B}
  \textbf{\bibinfo{volume}{86}}, \bibinfo{pages}{100506(R)}
  (\bibinfo{year}{2012}).

\bibitem[{\citenamefont{Weissman}(1988)}]{weissman1988RMP}
\bibinfo{author}{\bibfnamefont{M.~B.} \bibnamefont{Weissman}},
  \bibinfo{journal}{Rev. Mod. Phys.} \textbf{\bibinfo{volume}{60}},
  \bibinfo{pages}{537} (\bibinfo{year}{1988}).

\bibitem[{\citenamefont{Paladino et~al.}(2002)\citenamefont{Paladino, Faoro,
  Falci, and Fazio}}]{paladino2002PRL}
\bibinfo{author}{\bibfnamefont{E.}~\bibnamefont{Paladino}},
  \bibinfo{author}{\bibfnamefont{L.}~\bibnamefont{Faoro}},
  \bibinfo{author}{\bibfnamefont{G.}~\bibnamefont{Falci}}, \bibnamefont{and}
  \bibinfo{author}{\bibfnamefont{R.}~\bibnamefont{Fazio}},
  \bibinfo{journal}{Phys. Rev. Lett.} \textbf{\bibinfo{volume}{88}},
  \bibinfo{pages}{228304} (\bibinfo{year}{2002}).

\bibitem[{\citenamefont{D'Arrigo et~al.}(2005)\citenamefont{D'Arrigo, Flaci,
  Mastellone, and Paladino}}]{falci2005PhysE}
\bibinfo{author}{\bibfnamefont{A.}~\bibnamefont{D'Arrigo}},
  \bibinfo{author}{\bibfnamefont{G.}~\bibnamefont{Flaci}},
  \bibinfo{author}{\bibfnamefont{A.}~\bibnamefont{Mastellone}},
  \bibnamefont{and} \bibinfo{author}{\bibfnamefont{E.}~\bibnamefont{Paladino}},
  \bibinfo{journal}{Physica E} \textbf{\bibinfo{volume}{29}},
  \bibinfo{pages}{297} (\bibinfo{year}{2005}).

\bibitem[{\citenamefont{Zhou et~al.}(2010)\citenamefont{Zhou, Lang, and
  Joynt}}]{zhou2010QIP}
\bibinfo{author}{\bibfnamefont{D.}~\bibnamefont{Zhou}},
  \bibinfo{author}{\bibfnamefont{A.}~\bibnamefont{Lang}}, \bibnamefont{and}
  \bibinfo{author}{\bibfnamefont{R.}~\bibnamefont{Joynt}},
  \bibinfo{journal}{Quantum Inf. Process.} \textbf{\bibinfo{volume}{9}},
  \bibinfo{pages}{727} (\bibinfo{year}{2010}).

\bibitem[{\citenamefont{{Lo Franco} et~al.}(2012)\citenamefont{{Lo Franco},
  {D'Arrigo}, Falci, Compagno, and Paladino}}]{lofranco2012PhysScripta}
\bibinfo{author}{\bibfnamefont{R.}~\bibnamefont{{Lo Franco}}},
  \bibinfo{author}{\bibfnamefont{A.}~\bibnamefont{{D'Arrigo}}},
  \bibinfo{author}{\bibfnamefont{G.}~\bibnamefont{Falci}},
  \bibinfo{author}{\bibfnamefont{G.}~\bibnamefont{Compagno}}, \bibnamefont{and}
  \bibinfo{author}{\bibfnamefont{E.}~\bibnamefont{Paladino}},
  \bibinfo{journal}{Phys. Scripta} \textbf{\bibinfo{volume}{T147}},
  \bibinfo{pages}{014019} (\bibinfo{year}{2012}).

\bibitem[{\citenamefont{Galperin et~al.}(2006)\citenamefont{Galperin,
  Altshuler, Bergli, and Shantsev}}]{galperin2006PRL}
\bibinfo{author}{\bibfnamefont{Y.~M.} \bibnamefont{Galperin}},
  \bibinfo{author}{\bibfnamefont{B.~L.} \bibnamefont{Altshuler}},
  \bibinfo{author}{\bibfnamefont{J.}~\bibnamefont{Bergli}}, \bibnamefont{and}
  \bibinfo{author}{\bibfnamefont{D.~V.} \bibnamefont{Shantsev}},
  \bibinfo{journal}{Phys. Rev. Lett.} \textbf{\bibinfo{volume}{96}},
  \bibinfo{pages}{097009} (\bibinfo{year}{2006}).

\bibitem[{\citenamefont{Falci et~al.}(2005)\citenamefont{Falci, {D'Arrigo},
  Mastellone, and Paladino}}]{falci2005PRL}
\bibinfo{author}{\bibfnamefont{G.}~\bibnamefont{Falci}},
  \bibinfo{author}{\bibfnamefont{A.}~\bibnamefont{{D'Arrigo}}},
  \bibinfo{author}{\bibfnamefont{A.}~\bibnamefont{Mastellone}},
  \bibnamefont{and} \bibinfo{author}{\bibfnamefont{E.}~\bibnamefont{Paladino}},
  \bibinfo{journal}{Phys. Rev. Lett.} \textbf{\bibinfo{volume}{94}},
  \bibinfo{pages}{167002} (\bibinfo{year}{2005}).

\bibitem[{\citenamefont{Bellomo
  et~al.}(2010{\natexlab{b}})\citenamefont{Bellomo, Compagno, D'Arrigo, Falci,
  {Lo Franco}, and Paladino}}]{palermocatania2010PRA}
\bibinfo{author}{\bibfnamefont{B.}~\bibnamefont{Bellomo}},
  \bibinfo{author}{\bibfnamefont{G.}~\bibnamefont{Compagno}},
  \bibinfo{author}{\bibfnamefont{A.}~\bibnamefont{D'Arrigo}},
  \bibinfo{author}{\bibfnamefont{G.}~\bibnamefont{Falci}},
  \bibinfo{author}{\bibfnamefont{R.}~\bibnamefont{{Lo Franco}}},
  \bibnamefont{and} \bibinfo{author}{\bibfnamefont{E.}~\bibnamefont{Paladino}},
  \bibinfo{journal}{Phys. Rev. A} \textbf{\bibinfo{volume}{81}},
  \bibinfo{pages}{062309} (\bibinfo{year}{2010}{\natexlab{b}}).

\bibitem[{\citenamefont{Pasini and Uhrig}(2010)}]{pasini2010PRA}
\bibinfo{author}{\bibfnamefont{S.}~\bibnamefont{Pasini}} \bibnamefont{and}
  \bibinfo{author}{\bibfnamefont{G.~S.} \bibnamefont{Uhrig}},
  \bibinfo{journal}{Phys. Rev. A} \textbf{\bibinfo{volume}{81}},
  \bibinfo{pages}{012309} (\bibinfo{year}{2010}).

\bibitem[{\citenamefont{Wang and Liu}(2013)}]{liu2013PRA}
\bibinfo{author}{\bibfnamefont{Z.-Y.} \bibnamefont{Wang}} \bibnamefont{and}
  \bibinfo{author}{\bibfnamefont{R.-B.} \bibnamefont{Liu}},
  \bibinfo{journal}{Phys. Rev. A} \textbf{\bibinfo{volume}{87}},
  \bibinfo{pages}{042319} (\bibinfo{year}{2013}).

\bibitem[{\citenamefont{Nielsen and Chuang}(2000)}]{nielsenchuang}
\bibinfo{author}{\bibfnamefont{M.~A.} \bibnamefont{Nielsen}} \bibnamefont{and}
  \bibinfo{author}{\bibfnamefont{I.~L.} \bibnamefont{Chuang}},
  \emph{\bibinfo{title}{Quantum Computation and Quantum Information}}
  (\bibinfo{publisher}{Cambridge, UK, Cambridge University Press}, \bibinfo{year}{2000}).

\bibitem[{\citenamefont{D'Arrigo et~al.}()\citenamefont{D'Arrigo, {Lo Franco},
  Benenti, Paladino, and Falci}}]{darrigo2012arxiv}
\bibinfo{author}{\bibfnamefont{A.}~\bibnamefont{D'Arrigo}},
  \bibinfo{author}{\bibfnamefont{R.}~\bibnamefont{{Lo Franco}}},
  \bibinfo{author}{\bibfnamefont{G.}~\bibnamefont{Benenti}},
  \bibinfo{author}{\bibfnamefont{E.}~\bibnamefont{Paladino}}, \bibnamefont{and}
  \bibinfo{author}{\bibfnamefont{G.}~\bibnamefont{Falci}},
  \bibinfo{note}{arXiv:1207.3294}.

\bibitem[{\citenamefont{D'Arrigo et~al.}(2013)\citenamefont{D'Arrigo, {Lo
  Franco}, Benenti, Paladino, and Falci}}]{darrigo2013physscripta}
\bibinfo{author}{\bibfnamefont{A.}~\bibnamefont{D'Arrigo}},
  \bibinfo{author}{\bibfnamefont{R.}~\bibnamefont{{Lo Franco}}},
  \bibinfo{author}{\bibfnamefont{G.}~\bibnamefont{Benenti}},
  \bibinfo{author}{\bibfnamefont{E.}~\bibnamefont{Paladino}}, \bibnamefont{and}
  \bibinfo{author}{\bibfnamefont{G.}~\bibnamefont{Falci}},
  \bibinfo{journal}{Phys. Scripta} \textbf{\bibinfo{volume}{T153}},
  \bibinfo{pages}{014014} (\bibinfo{year}{2013}).

\bibitem[{\citenamefont{{D'Arrigo} et~al.}(2014)\citenamefont{{D'Arrigo},
  Benenti, {Lo Franco}, Falci, and Paladino}}]{darrigo2014}
\bibinfo{author}{\bibfnamefont{A.}~\bibnamefont{{D'Arrigo}}},
  \bibinfo{author}{\bibfnamefont{G.}~\bibnamefont{Benenti}},
  \bibinfo{author}{\bibfnamefont{R.}~\bibnamefont{{Lo Franco}}},
  \bibinfo{author}{\bibfnamefont{G.}~\bibnamefont{Falci}}, \bibnamefont{and}
  \bibinfo{author}{\bibfnamefont{E.}~\bibnamefont{Paladino}},
  \bibinfo{journal}{Int. J. Quant. Inf.} \textbf{\bibinfo{volume}{12}},
  \bibinfo{pages}{1461005} (\bibinfo{year}{2014}).

\bibitem[{\citenamefont{M\"ott\"onen et~al.}(2006)\citenamefont{M\"ott\"onen,
  de~Sousa, Zhang, and Whaley}}]{Mottonen2006}
\bibinfo{author}{\bibfnamefont{M.}~\bibnamefont{M\"ott\"onen}},
  \bibinfo{author}{\bibfnamefont{R.}~\bibnamefont{de~Sousa}},
  \bibinfo{author}{\bibfnamefont{J.}~\bibnamefont{Zhang}}, \bibnamefont{and}
  \bibinfo{author}{\bibfnamefont{K.~B.} \bibnamefont{Whaley}},
  \bibinfo{journal}{Phys. Rev. A} \textbf{\bibinfo{volume}{73}},
   \bibinfo{pages}{022332} (\bibinfo{year}{2006}).

\bibitem[{\citenamefont{Gordon et~al.}(2008)\citenamefont{Gordon, Kurizki, and
  Lidar}}]{Gordon2008}
\bibinfo{author}{\bibfnamefont{G.}~\bibnamefont{Gordon}},
  \bibinfo{author}{\bibfnamefont{G.}~\bibnamefont{Kurizki}}, \bibnamefont{and}
  \bibinfo{author}{\bibfnamefont{D.~A.} \bibnamefont{Lidar}},
  \bibinfo{journal}{Phys. Rev. Lett.} \textbf{\bibinfo{volume}{101}},
   \bibinfo{pages}{010403} (\bibinfo{year}{2008}).

\bibitem[{\citenamefont{Gorman et~al.}(2012)\citenamefont{Gorman, Young, and
  Whaley}}]{Gorman2012}
\bibinfo{author}{\bibfnamefont{D.~J.} \bibnamefont{Gorman}},
  \bibinfo{author}{\bibfnamefont{K.~C.} \bibnamefont{Young}}, \bibnamefont{and}
  \bibinfo{author}{\bibfnamefont{K.~B.} \bibnamefont{Whaley}},
  \bibinfo{journal}{Phys. Rev. A} \textbf{\bibinfo{volume}{86}},
  \bibinfo{pages}{012317} (\bibinfo{year}{2012}).

\bibitem[{\citenamefont{{D'Arrigo} et~al.}(2008)\citenamefont{{D'Arrigo},
  Falci, Mastellone, and Paladino}}]{ka:208-darrigo-njp-corrnoise}
\bibinfo{author}{\bibfnamefont{A.}~\bibnamefont{{D'Arrigo}}},
  \bibinfo{author}{\bibfnamefont{G.}~\bibnamefont{Falci}},
  \bibinfo{author}{\bibfnamefont{A.}~\bibnamefont{Mastellone}},
  \bibnamefont{and} \bibinfo{author}{\bibfnamefont{E.}~\bibnamefont{Paladino}},
  \bibinfo{journal}{New J. Phys.} \textbf{\bibinfo{volume}{10}},
  \bibinfo{pages}{115006} (\bibinfo{year}{2008}).

\end{thebibliography}
\end{document}